\documentclass[aps,prd,12pt,superscriptaddress,nofootinbib]{revtex4-2}
 \usepackage{bm}
 \usepackage{amsmath}
 \usepackage{graphicx}
 \usepackage{amssymb}
 \usepackage{subfigure}
 \usepackage{amssymb}
 \usepackage{hyperref}
 \usepackage{CJK}
 \hypersetup{
 	colorlinks=true,
 	linkcolor=red,
 	citecolor=blue,
 }
 \usepackage[utf8]{inputenc}
 \hypersetup{
 	colorlinks=true,
 	linkcolor=red,
 	citecolor=blue,
 }
 \begin{document}
 \begin{CJK*}{UTF8}{gbsn}
 	\title{Primordial black holes and scalar-induced secondary gravitational waves from inflationary models with a noncanonical kinetic term}
 	\author{Zhu Yi (易竹)}
 	\email{yz@bnu.edu.cn}
 	\affiliation{Department of Astronomy, Beijing Normal University, Beijing 100875, China}

 	\author{Qing Gao (郜青)}
    \email{gaoqing1024@swu.edu.cn}
    \affiliation{School of Physical Science and Technology, Southwest University, Chongqing 400715, China}

 	\author{Yungui Gong (龚云贵)}
 	\email{Corresponding author. yggong@hust.edu.cn}
 	\affiliation{School of Physics, Huazhong University of Science and Technology, Wuhan, Hubei 430074, China}

   	\author{Zong-hong Zhu (朱宗宏)}
 	\email{zhuzh@bnu.edu.cn}
 	\affiliation{Department of Astronomy, Beijing Normal University, Beijing 100875, China}

\begin{abstract}
With the enhancement mechanism provided by a noncanonical kinetic term with a peak,
the amplitude of primordial curvature perturbations can be enhanced by seven orders of magnitude at small scales while keeping to be consistent with observations at large scales.
The peak function and inflationary potential are not restricted in this mechanism.
We use the Higgs model and T-model as examples to show how abundant primordial black hole dark matter with
different mass and scalar induced secondary gravitational waves with different peak frequency are generated.
We also show that the enhanced power spectrum for the primordial curvature perturbations and the energy density of the scalar induced secondary gravitational waves can have either a sharp peak or a broad peak.
The primordial black holes with the mass around $10^{-14}-10^{-12} M_{\odot}$ produced with the enhancement mechanism can make up almost all dark matter,
and the scalar induced secondary gravitational waves accompanied with the production of primordial black holes can be tested by the pulsar timing arrays and spaced based gravitational wave observatory.
Therefore, the mechanism can be tested by primordial black hole dark matter and gravitational wave observations.
\end{abstract}

\preprint{2011.10606}

\maketitle

\end{CJK*}

\section{Introduction}	
The detection of gravitational wave (GW)   by the Laser  Interferometer Gravitational Wave Observatory
(LIGO) Scientific  Collaboration and the Virgo Collaboration announced the dawn of the  era of multimessenger astronomy  \cite{Abbott:2016blz,Abbott:2016nmj, Abbott:2017vtc,Abbott:2017oio,TheLIGOScientific:2017qsa, Abbott:2017gyy,LIGOScientific:2018mvr,Abbott:2020uma,LIGOScientific:2020stg,Abbott:2020khf,Abbott:2020tfl,Abbott:2020niy}.
It was pointed out that these GWs may be emitted by the mergers of the  stellar mass primordial black holes (PBHs)   \cite{Bird:2016dcv,Sasaki:2016jop}.
PBHs are also proposed to account for dark matter (DM) due to the failure of direct detection of the particle DM \cite{Ivanov:1994pa,Frampton:2010sw,Belotsky:2014kca,Khlopov:2004sc,Clesse:2015wea,Carr:2016drx,Inomata:2017okj,Garcia-Bellido:2017fdg,Kovetz:2017rvv,Carr:2020xqk},
and PBHs with the masses around $10^{-17}-10^{-15} M_{\odot}$ and $10^{-14}-10^{-12}M_{\odot}$ can make up almost all of DM
because the abundances of PBHs at these mass windows are not constrained by observations.
Furthermore, PBHs with the mass several times of the Earth's  can explain the Planet 9
which is a hypothetical astrophysical object in the outer solar system used to explain  the anomalous orbits of trans-Neptunian objects \cite{Scholtz:2019csj}.
PBHs are formed from the gravitational collapse of overdense regions with their density contrasts at the horizon reentry during radiation domination exceeding the threshold value \cite{Carr:1974nx,Hawking:1971ei}.
The overdense  regions may be seeded from the primordial curvature  perturbations generated during inflation.
To produce enough abundance of PBH DM,
the amplitude of the power spectrum  of the primordial curvature perturbations  should be $A_s\sim \mathcal{O}(0.01)$, while the  constraint on the  amplitude of the power spectrum  at   large scales  from the  cosmic microwave background (CMB) anisotropy measurements is $A_s=2.1\times 10^{-9}$ \cite{Akrami:2018odb}.
Therefore, the feasible way to produce enough abundance of  PBH DM is by enhancing the amplitude of the power spectrum  at least seven orders of magnitude to reach the threshold at small scales \cite{Gong:2017qlj,Lu:2019sti,Sato-Polito:2019hws}.

An effective way  to enhance the power spectrum  is through the ultra-slow-roll inflation \cite{Martin:2012pe,Motohashi:2014ppa,Yi:2017mxs}.
For the single field inflation with a canonical scalar field,
introducing  an inflection point in  the potential  is a very economic way to realize  the ultra-slow-roll inflation,
hence producing abundant PBH DM \cite{Garcia-Bellido:2017mdw,Germani:2017bcs,Motohashi:2017kbs,Ezquiaga:2017fvi,Gong:2017qlj,Ballesteros:2018wlw,Dalianis:2018frf}.
However, it is not an easy task to achieve the big enhancement on
the power spectrum while keeping the total number of e-folds
around $50-60$ by fine tuning the model parameters \cite{Sasaki:2018dmp,Passaglia:2018ixg}.
Nonminimal coupling to gravity and noncanonical kinetic terms were then considered \cite{Kamenshchik:2018sig,Fu:2019ttf,Fu:2019vqc,Dalianis:2019vit,Lin:2020goi,Braglia:2020eai,Gundhi:2020zvb,Cheong:2019vzl}.
With the coupling parameter $1/M^2$ for the nonminimally derivative coupling to Einstein tensor
generalizing to  a special function $g(\phi)=h/\sqrt{1+(\phi-\phi_p)^2/w^2}$,
the inflationary model with the  potential $\phi^{2/5}$ succeeds enhancing the power spectrum up to seven orders of magnitude at small scales \cite{Fu:2019ttf,Fu:2019vqc,Dalianis:2019vit},
but both the potential and the coupling function $g(\phi)$ in this mechanism are restricted
to the specific forms, and we need to fine tune the model parameters.
Motivated by k inflation \cite{ArmendarizPicon:1999rj,Garriga:1999vw}
and G inflation \cite{Kobayashi:2010cm,Kobayashi:2011nu,Kobayashi:2011pc,Herrera:2018ker},
the noncanonical kinetic term $[1+G_p(\phi)]\dot\phi^2/2$ with the peak function $G_p=h/(1+|\phi-\phi_p|/w)$
was proposed to enhance the power spectrum and produce abundant PBH DM \cite{Lin:2020goi}.
Although the problem with the fine tuning of the model parameters is eased in this mechanism,
but the  potential was restricted to be $\phi^{2/5}$ \cite{Lin:2020goi} or $\phi^{1/3}$ \cite{Gao:2021vxb}.
This enhancement mechanism was then improved by generalizing the noncanonical
kinetic term to be  $G_p(\phi)+f(\phi)$ \cite{Yi:2020kmq}.
The function $G_p$ is used to enhance the power spectrum  and the function $f(\phi)$ is acted as a chameleon field to modify the shape of the potential during inflation to make the model  consistent with the observations.
In this improved mechanism, it was shown that the inflation driven by the Higgs bonson with the potential $\lambda \phi^4/4$ in the standard model of particle physics
satisfies the CMB constraints and provides the seed of PBH DM \cite{Yi:2020kmq}.  
The improved mechanism also works for natural inflation \cite{Gao:2020tsa} and T-model, and other peak functions are also possible,
so the potential for the inflaton and the peak function are not restricted
in this mechanism.

Accompanied by the formation of PBHs, the large scalar perturbations at small scales induce secondary  GWs after the horizon reentry during  the radiation dominated epoch \cite{Matarrese:1997ay,Mollerach:2003nq,Ananda:2006af,Baumann:2007zm,Garcia-Bellido:2017aan,Saito:2008jc,Saito:2009jt,Bugaev:2009zh,Bugaev:2010bb,Alabidi:2012ex,Orlofsky:2016vbd,Nakama:2016gzw,Inomata:2016rbd,Cheng:2018yyr,Cai:2018dig,Bartolo:2018rku,Bartolo:2018evs,Kohri:2018awv,Espinosa:2018eve,Cai:2019amo,Cai:2019elf,Cai:2019bmk,Cai:2020fnq,Domenech:2019quo,Domenech:2020kqm,Fumagalli:2020adf,Fumagalli:2020nvq,Pi:2020otn}.
These scalar induced gravitational waves (SIGWs) that have a vast range of frequencies  and  consist  of  the  stochastic  background can be detected by pulsar timing arrays (PTA) \cite{Ferdman:2010xq,Hobbs:2009yy,McLaughlin:2013ira,Hobbs:2013aka,Moore:2014lga}
and the space based GW detectors such as Laser Interferometer Space Antenna (LISA) \cite{Danzmann:1997hm,Audley:2017drz}, Taiji \cite{Hu:2017mde} and TianQin  \cite{Luo:2015ght} in the future,
and  in turn disclose the properties of PBHs and primordial power spectrum at small scales.

In this paper, we elaborate on the mechanism proposed in Ref. \cite{Yi:2020kmq} in detail.
We consider a general peak function  $G_p(\phi)=h/[1+\left(|\phi-\phi_p|/{w}\right)^q]$
and show that both sharp and broad peaks can be generated in this mechanism.
The scale where the power spectrum is enhanced can be adjusted by the parameter $\phi_p$,
and the peak shape of the power spectrum can be adjusted by the power index $q$.
The paper is organized as follows. In Sec. II, we review the enhancement mechanism for primordial curvature perturbations proposed in Ref. \cite{Yi:2020kmq}.
We discuss the production of PBH DM and SIGWs from the Higgs model in Sec. III,
and the generation of both PBH DM and SIGWs from T-model is  considered in Sec. IV.
We conclude the paper in Sec. V.

\section{The enhancement mechanism}
In this section, we review the enhancement mechanism of the power spectrum at small scales proposed in Refs. \cite{Lin:2020goi,Yi:2020kmq}.  The action is
\begin{equation}\label{act1}
  S=\int d x^4 \sqrt{-g}\left[\frac{1}{2}R+X+G(\phi)X-V(\phi)\right],
\end{equation}
where $X=-g_{\mu\nu}\nabla^{\mu}\phi\nabla^{\nu}\phi/2$,  the noncanonical kinetic term may arise from scalar-tensor theory of gravity,
G inflation \cite{Kobayashi:2010cm} or k inflation \cite{ArmendarizPicon:1999rj,Garriga:1999vw},
and we take the convention $8\pi G=1$. After inflation,
we expect $G(\phi)$ is negligible
so that the noncanonical kinetic term disappears.
The background equations are
\begin{gather}
\label{Eq:eom1}
3H^2=\frac{1}{2}\dot{\phi}^2+V(\phi)+\frac{1}{2}\dot{\phi}^2G(\phi),\\
\label{Eq:eom2}
\dot{H}=-\frac{1}{2}[1+G(\phi)]\dot{\phi}^2,\\
\label{Eq:eom3}
\ddot{\phi}+3H\dot{\phi}+\frac{V_{\phi}+\dot{\phi}^2G_{\phi}/2}{1+G(\phi)}=0,
\end{gather}
where $G_\phi=dG(\phi)/d\phi$ and $V_\phi=dV/d\phi$.  The slow-roll parameters are defined as
\begin{equation}\label{sl}
\epsilon_H=-\frac{\dot{H}}{H^2},\quad  \eta_H=-\frac{\ddot{\phi}}{H\dot{\phi}},\quad
 \epsilon_G=-\frac{G_{\phi}\dot{\phi}^2}{2V_{\phi}},
\end{equation}
and the corresponding  slow-roll conditions are
\begin{equation}\label{slcon}
| \epsilon_H|\ll 1,\quad |\eta_H|\ll1,\quad |\epsilon_G|\ll1.
\end{equation}

Under these slow-roll conditions, the background Eqs. \eqref{Eq:eom1} and \eqref{Eq:eom3} become
\begin{gather}
\label{abg1}
  3H^2\approx V, \\
  \label{abg2}
  3H\dot{\phi}(1+G)+V_{\phi}\approx 0.
\end{gather}
Combining the slow-roll background equations and the definitions of the slow-roll parameters, we have
\begin{gather}
    \label{slv1}
    \epsilon_H\approx \frac{\epsilon_V}{1+G},\\
    \label{slv2}
    \eta_H\approx \frac{\eta_V}{1+G}-\frac{\epsilon_V}{1+G}-\frac{\sqrt{2\epsilon_V}G_{\phi}}{(1+G)^2},
\end{gather}
where the potential slow-roll parameters are defined as
\begin{equation}\label{slp}
  \epsilon_V=\frac{1}{2}\left(\frac{V_{\phi}}{V}\right)^2, \quad \eta_V=\frac{V_{\phi\phi}}{V}.
\end{equation}

The power spectrum  for the curvature perturbation is
\begin{equation}
\label{ps}
\mathcal{P}_\zeta=\frac{H^4}{4\pi^2\dot{\phi}^2(1+G)}\approx \frac{V^3}{12\pi^2V_{\phi}^2}(1+G).
\end{equation}
If the noncanonical kinetic term disappears, $G=0$, then
we recover the result in standard slow-roll inflation.
In other words, the noncanonical kinetic coupling $G$ can be used
to enhance the scalar power spectrum at small scales.
The scalar spectral index is
\begin{equation}
\label{nseq10}
\begin{split}
  n_s-1&=\frac{d\ln \mathcal{P}_\zeta}{d \ln k}=-4\epsilon_H+2\eta_H-\frac{G_\phi}{1+G}\frac{\dot{\phi}}{H}\\
  &\approx \frac{1}{1+G}\left(2\eta_V-6\epsilon_V-
\frac{G_{\phi}}{1+G}\sqrt{2\epsilon_V}\right).
  \end{split}
\end{equation}

Because tensor perturbations are independent of scalar perturbations and the additional  noncanonical kinetic term doesn't introduce any tensor degree of freedom, so the tensor power spectrum is the same as that in the canonical case
\begin{equation}\label{pt}
\mathcal{P}_T=\frac{2 H^2}{\pi^2}\approx \frac{2V}{3\pi^2}.
\end{equation}
Combining the scalar power spectrum \eqref{ps} and the tensor power spectrum \eqref{pt}, we obtain the tensor-to-scalar ratio
\begin{equation}
\label{req1}
r \simeq \frac{16\epsilon_V}{1+G}.
\end{equation}

Since the Planck 2018 results give $A_s=2.1\times 10^{-9}$
at the pivotal scale $k_*=0.05$ Mpc$^{-1}$ \cite{Akrami:2018odb},
in order to produce enough abundance PBH DM, the scalar power spectrum should be enhanced at least seven orders of magnitude to reach $A_s\sim \mathcal{O}(0.01)$ at small scales.
From the power spectrum \eqref{ps}, we see that a big peak in the coupling function $G(\phi)$ could realize  this purpose.
If we want the effect of the peak function is to enhance the power spectrum at small scales only while keeping the predictions of $n_s$ and $r$ at large scales,
the peak function should be negligible away from the peak.
Inspired by the coupling $\omega(\phi)=1/\phi$ in Brans-Dicke theory \cite{Brans:1961sx},  a suitable peak function  is \cite{Lin:2020goi}
\begin{equation}\label{pfunc}
G_p(\phi)=\frac{h}{1+|\phi-\phi_p|/w},
\end{equation}
where $h$ gives the amplitude of the peak,  $w$  controls the width of the peak, and $\phi_p$ determines the position of the peak in the power spectrum. Because the peak function is  negligible at the CMB scale, from Eq. \eqref{nseq10} and Eq. \eqref{req1}, the scalar spectral index and the tensor-to-scalar ratio   reduce  to the standard canonical slow-roll inflation results
\begin{gather}
\label{nseqst}
n_s-1=2\eta_V-6\epsilon_V, \\
\label{reqst}
r \simeq 16\epsilon_V.
\end{gather}

With the help of Eqs. \eqref{abg1} and \eqref{abg2}, we obtain the number of
$e$-folds  before the end of inflation at the horizon exit  for the pivotal scale,
\begin{equation}\label{efold0}
N=\int_{\phi_e}^{\phi_*}\frac{V}{V_{\phi}}d \phi+\Delta N,
\end{equation}
where the first term is the $e$-folding number from the standard slow-roll inflation and the second term is from the peak function
\begin{equation}\label{efold:peak}
  \Delta N=\frac{V(\phi_p)}{V_{\phi}(\phi_p)}\int_{\phi_p+\Delta \phi}^{\phi_p-\Delta \phi}G d\phi.
\end{equation}
Although the peak function does not affect $n_s$ and $r$ at large scales,
it influences $e$-folding number and contributes to about $20$ $e$-folds,
henceforth effectively moves $\phi_*$ closer to $\phi_e$ in order to keep the total number of $e$-folds around $60$.
In summary, the predictions of $n_s$ and $r$ at the pivotal scale for the inflationary model with the noncanonical kinetic term and $e$-folding number around 60 is the same as the canonical one with the $e$-folding number around $40$.

The Planck 2018 constraints \cite{Akrami:2018odb,Ade:2018gkx}
\begin{equation}\label{cmb:con}
\begin{split}
n_s = 0.9649\pm 0.0042  ~(68\% \text{CL}) ,\\
r_{0.05} < 0.06 ~(95\% \text{CL}),
\end{split}
\end{equation}
favor the parametrization
\begin{equation}\label{ns:2}
  n_s=1-\frac{2}{N},
\end{equation}
with $N=60$. This result can be obtained from many inflationary models \cite{Starobinsky:1980te,Kaiser:1994vs,Bezrukov:2007ep}.
With the $e$-folding number reducing to $40$, the formula \eqref{ns:2} is inconsistent with  the observational data, and it should be modified   to
\begin{equation}\label{ns:43}
n_s=1-\frac{4}{3N},
\end{equation}
to be consistent with the observational data with $N=40$.
To get the inflationary potential with the  prediction \eqref{ns:43},
we can use the method of potential reconstruction \cite{ Lin:2015fqa}.
The reconstructed potential from the general parametrization
\begin{equation}\label{pre:ns1}
 n_s=1-\frac{n+2}{2N},
\end{equation}
is   the   chaotic inflation \cite{Linde:1983gd, Lin:2015fqa}
\begin{equation}\label{chaotic:p}
  V(\phi)=V_0 \phi^n,
\end{equation}
and the corresponding  tensor-to-scalar ratio is
\begin{equation}\label{pre:nsr}
r=\frac{4 n}{ N}.
\end{equation}
 Comparing Eq. \eqref{pre:ns1} to Eq. \eqref{ns:43},
 we get $n=2/3$,  and the predictions are
\begin{equation}\label{pre:23}
  n_s=0.967,\quad r=0.067,
\end{equation}
where the tensor-to-scalar ratio is a bit large compared with the observational constraints  \eqref{cmb:con}.
To obtain a smaller tensor-to-scalar ratio, from Eq. \eqref{pre:nsr}, we need a smaller $n$.  For the sake of simplicity, we   consider $n=1/3$, and the predictions are
\begin{equation}\label{pre:13}
  n_s=0.971,\quad r=0.033.
\end{equation}
In fact, the allowed values of $n$ from the observational constraints are broad, for example,  the potential with $n=2/5$ discussed in Ref. \cite{Lin:2020goi} gives
\begin{equation}
n_s=0.970,\quad r=0.040,
\end{equation}
which is consistent with the observational constraints.
On the other hand, the $e$-folding number contributed from the peak function is not  exact $20$,
it depends on the model parameters.
If the peak function contributes $25$ $e$-folds,
the effective number of $e$-folds is $35$,
from   Eq. \eqref{pre:ns1} and Eq. \eqref{pre:nsr}, the potential with $n=1/3$ and $N=35$ gives $n_s=0.967$ and $r=0.038$ which satisfy the observational constraints \eqref{cmb:con}.
Therefore, for suitable model parameters,
the  predictions of these models could be consistent with CMB constraints.

Because the effective $e$-folding number is about $40$, the usual inflationary potentials that satisfy the observational constraints with $N\sim60$ would become incompatible with CMB observations with the enhancement mechanism.
Therefore, the application of this mechanism is limited.
To overcome this limitation, we take the advantage of the noncanonical kinetic
term with a peak function and the success of the power-law potential $U(\Phi)=U_0\Phi^n$,  so we take the form of the noncanonical coupling function to be
\begin{equation}\label{gfunc1}
 G(\phi)=G_p(\phi)+f(\phi),
\end{equation}
where the peak function $G_p(\phi)$  is used to enhance the scalar power spectrum at small scales,
the function $f(\phi)$ is acted as a chameleon field so that the potential $V(\phi)$
is adjusted to become the power-law potential $U(\Phi)$.
More specifically, around the peak $f(\phi)\ll G_p(\phi)$, the peak function
enhances the scalar power spectrum and contributes about 20 $e$-folds.
Away from the peak we use the
function $f(\phi)$ to change the noncanonical scalar field with the potential $V(\phi)$
to the canonical scalar field $\Phi$ with the potential $U(\Phi)$.
In particular, under  the transformation
\begin{equation}\label{trans}
  d\Phi=\sqrt{1+f(\phi)}d\phi, \quad U(\Phi)=V[\phi(\Phi)],
\end{equation}
the action \eqref{act1} becomes
\begin{equation}\label{act2}
  S=\int d x^4 \sqrt{-g}\left[\frac{1}{2}R-\frac{1}{2}g_{\mu\nu}\nabla^\mu\Phi\nabla^\nu\Phi-U(\Phi)\right].
\end{equation}
With the help of the chameleon function $f(\phi)$, we arrive at the standard case with the canonical scalar field $\Phi$, so
at large scales the scalar spectral index and the tensor-to-scalar ratio  are
\begin{gather}
\label{nseq3}
n_s-1\simeq  2\eta_U-6\epsilon_U, \\
\label{req3}
r \simeq 16\epsilon_U,
\end{gather}
where $\epsilon_U$ and $\eta_U$ are
\begin{equation}\label{slp2}
  \epsilon_U=\frac{1}{2}\left(\frac{U_{\Phi}}{U}\right)^2, \quad \eta_U=\frac{U_{\Phi\Phi}}{U}.
\end{equation}
Therefore, the effect of the function $f(\phi)$ is to reshape the potential $V(\phi)$ to $U(\Phi)$ so that the predictions could be consistent with the observational constraints.
In other words, for a given potential $V(\phi)$, from the transformation \eqref{trans}, we can find a corresponding  function $f(\phi)$ and keep   the predictions for  $n_s$ and $r$ the same as   that  given by the effective potential   $U(\Phi)$.
For example, for the power-law potential $U(\Phi)=U_0 \Phi^n$, the relation between $f(\phi)$ and potential $V(\phi)$ is
\begin{gather}
\label{gfuncn}
f(\phi)= \frac{1}{n^2}\left(\frac{1}{U_0}\right)^{2/n}V^{\frac{2}{n}-2}V_\phi^2.
\end{gather}
The chaotic inflation with $n=1/3$ and $N\approx 40$ is consistent with the observational constraints as discussed above,
so the potential $V(\phi)$  is not restricted to be a specific form.

\begin{figure}[htbp]
  \centering
  \includegraphics[width=0.9\columnwidth]{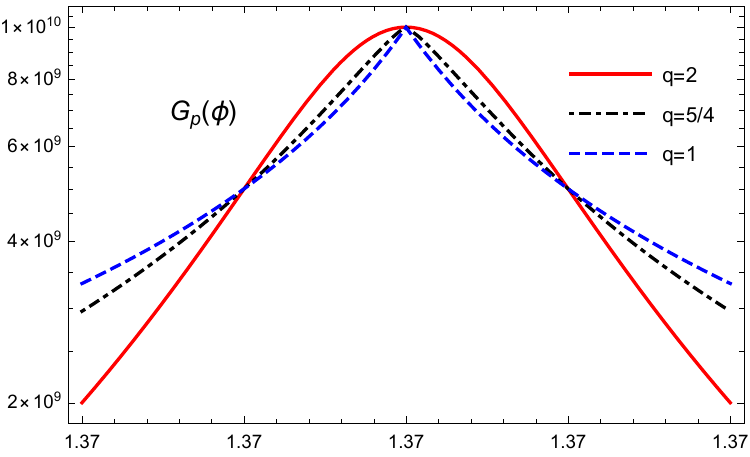}
  \caption{The peak function $G_p(\phi)$ with different $q$.  }\label{gpic}
\end{figure}

Besides the loose restriction on the   potential,   to enhance the power spectrum, the peak function  $G_p(\phi)$ is not restricted to the form  \eqref{pfunc} too.
The peak function with a similar form as Eq. \eqref{pfunc} can also successfully enhance the scalar power spectrum to produce  abundant  PBH  DM  and contribute about 20 $e$-folds,
such as the peak function
\begin{equation}
    G_p=\frac{h}{\sqrt{1+\left(\frac{\phi-\phi_p}{w}\right)^2}},
\end{equation}
which is discussed in Refs. \cite{Fu:2019ttf,Lin:2020goi}.  In this paper, we consider a general peak function
\begin{equation}\label{gfuncng}
  G_p(\phi)=\frac{h}{1+\left(|\phi-\phi_p|/{w}\right)^q},
\end{equation}
where the power index $q$  controls the shape of the enhanced power spectrum.  The peak function with a larger $q$ has a flatter peak,
leading to  a  wider peak in the  power spectrum.
Therefore,  by choosing different $q$ in the peak function,
we can obtain different shapes for the enhanced power spectrum.
The peak function with different $q$ is  shown in Fig. \ref{gpic}.

In summary, in our  mechanism, the predictions for $n_s$ and $r$ are determined by the effective potential $U(\Phi)$
and  the original  potential in the action \eqref{act1} is not restricted to a special form
because the function $f(\phi)$ can reshape it to the effective potential;
the shape of the enhanced power spectrum is controlled by $q$ in the  peak function.
In the next sections, we apply this mechanism  to Higgs model and T-model.

\section{Higgs model}

In this section, we show that the Higgs field can not only drive inflation but also  explain DM in our universe in terms of PBH.  The Higgs potential is
\begin{equation}\label{higgs:potential}
V(\phi)=\frac{\lambda}{4}\phi^4,
\end{equation}
where the coupling constant $\lambda\approx 0.13$ \cite{Patrignani:2016xqp}.
To obtain the predictions consistent with the observational constraints,
we consider the effective potential $U(\phi)=U_0\Phi^n$ with $n=1/3$ and $n=2/3$,
other values of $n$ such as $n=2/5$ and other forms of $U(\Phi)$ are also possible.
Substituting the Higgs potential \eqref{higgs:potential} into Eq. \eqref{gfuncn},
we obtain
\begin{equation}\label{higgs:fn}
  f(\phi)=f_0 \phi^{-2+8/n},
\end{equation}
where
\begin{equation}
f_0=\frac{16}{n^2}\left(\frac{\lambda}{4U_0}\right)^{2/n}.
\end{equation}
For $n=1/3$,  we get $f(\phi)=f_0\phi^{22}$.
For $n=2/3$,  we get $f(\phi)=f_0\phi^{10}$.
At low energy scales $\phi\ll 1$, $f(\phi)$ is negligible and the model reduces to
the standard case with the canonical kinetic term.

As mentioned above, the peak shape of the enhanced power spectrum is determined by $q$ in the peak function \eqref{gfuncng}.
To obtain different shapes of peak in the power spectrum,
we consider the peak function \eqref{gfuncng} with $q=1$ and $q=5/4$ to demonstrate the sharp and broad peaks.
To  distinguish different models, as shown in Table \ref{higglabel},
we label the model with $n=1/3$ and $q=1$ as H11, the model with $n=1/3$ and $q=5/4$ as H12,
the model with $n=2/3$ and $q=1$ as H21, and the model with $n=2/3$ and $q=5/4$ as H22, respectively.
\begin{table}[htbp]
	\begin{tabular}{lll}
		\hline
		\hline
		Label \quad   &  $U(\phi)=U_0\Phi^n$ \quad\quad& $q$ \\
		\hline
 		H11 \quad       & $n=1/3$  & $q=1$\\
 		H12 \quad      & $n=1/3$  & $q=5/4$\\
 		H21 \quad      & $n=2/3$  & $q=1$\\
 		H22 \quad    & $n=2/3$ & $q=5/4$\\
		\hline
		\hline
	\end{tabular}
	\caption{The labeling for the Higgs models with different peak functions and effective potentials.}
\label{higglabel}
\end{table}

\begin{figure}[htbp]
  \centering
  \includegraphics[width=0.95\columnwidth]{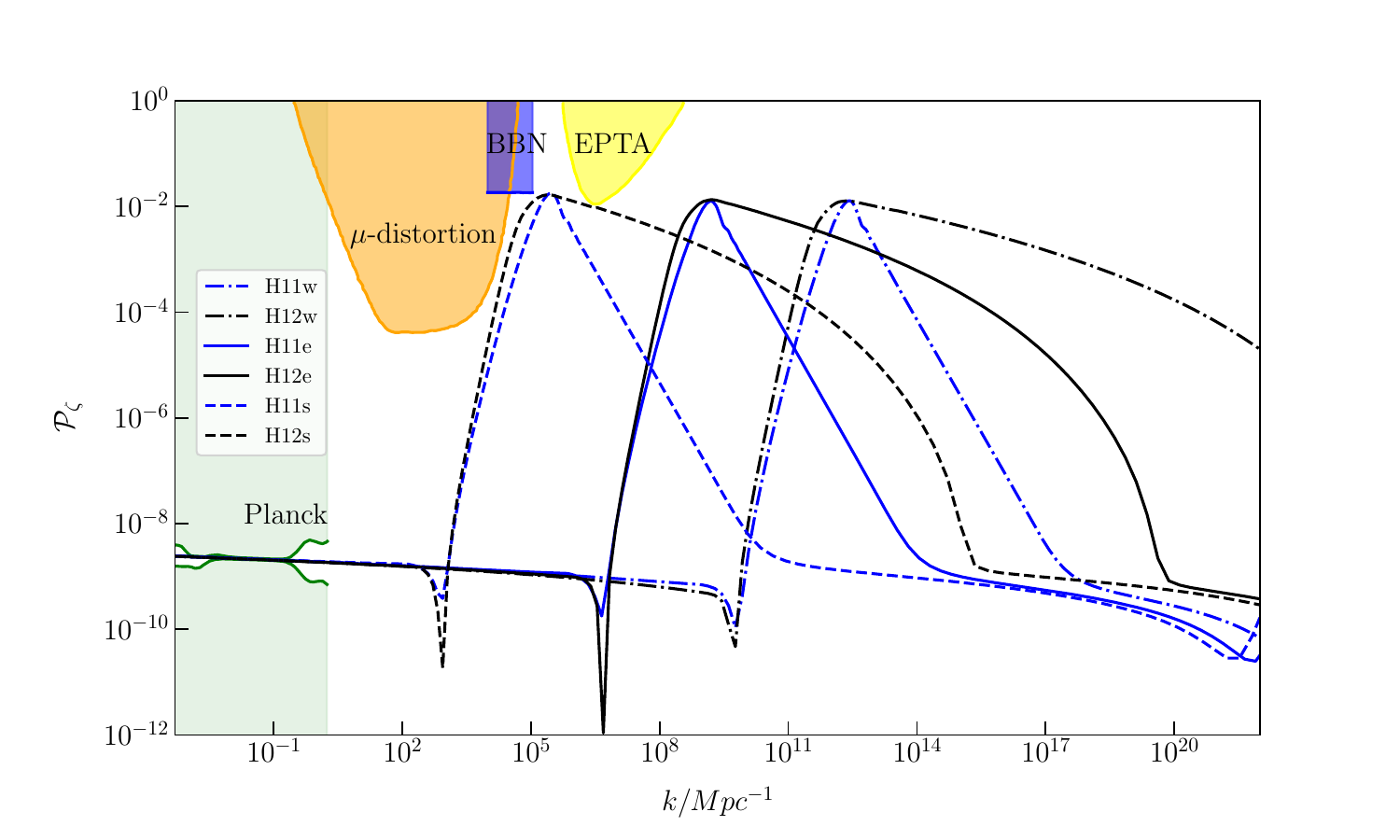}
  \caption{The enhanced power spectrum in the Higgs model. The blue and black lines
  denote the results for the models H11 with sharp peak and H12 with broad peak, respectively. The light green shaded region is excluded by the CMB observations \cite{Akrami:2018odb}. The yellow, blue and orange regions show the
constraints from the PTA observations \cite{Inomata:2018epa},
the effect on the ratio between neutron and proton
during the big bang nucleosynthesis (BBN) \cite{Inomata:2016uip}
and $\mu$-distortion of CMB \cite{Fixsen:1996nj}, respectively.}
  \label{pr:higgs}
\end{figure}

Choosing the value $\phi_*$ of the scalar field  at the pivotal scale,
the parameters $f_0$, $h$, $w$ and $\phi_p$ in the peak function as shown in Table \ref{higgst1},
we numerically solve the background equations \eqref{Eq:eom1}-\eqref{Eq:eom3} and the perturbation equation
\begin{equation}
\label{zeta:k}
\frac{d^2u_k}{d\eta^2}+\left(k^2-\frac{1}{z}\frac{d^2z}{d\eta^2}\right)u_k=0,
\end{equation}
to obtain the scalar power spectrum $\mathcal{P}_\zeta=k^3|\zeta_k|^2/(2\pi^2)$,
where the conformal time $\eta=\int dt/a(t)$, $u_k=z\zeta_k$
and $z=a\dot{\phi}(1+G)^{1/2}/H$.
In Fig. \ref{pr:higgs}, we show the results of the power spectra for the models H11 and H12 with blue and black lines respectively.
The results for the models H21 and H22 are similar,
so we don't show them in the figures.
From Fig. \ref{pr:higgs},
we see that the power spectrum produced in the model H11 has a sharp peak
while the power spectrum produced in the model H12 has a broad peak.
Therefore, we can adjust the peak shape by choosing $q$. Additionally,
the peak position or the scale where the power spectrum is enhanced
can be adjusted by the parameter $\phi_p$
as shown in Table \ref{higgst1} and Fig. \ref{pr:higgs}.
To distinguish the scale for the enhanced power spectrum, we use
additional labels ``w," ``e," and ``s" as shown in Table \ref{higgst1}.
We also numerically calculate the scalar spectral index $n_s$
and the tensor-to-scalar ratio $r$ for these models and the
results are shown in Table \ref{higgst1}.
As expected, the peak function has little influence on $n_s$ and $r$,
and $n_s$ and $r$ are determined by the effective potential.
For the models H11 and H12, we get
\begin{equation}\label{higgs:phi13}
  n_s\approx 0.967, \quad r\approx0.040.
\end{equation}
These results are consistent with the observational constraints
and the slow-roll predictions \eqref{pre:13}.
If $f_0$ is fixed as $f_0=1$ and the coupling constant $\lambda$ is chosen as $\lambda\sim 10^{-9}$,
then similar power spectrum can be obtained \cite{Yi:2020kmq}.
When the running of Higgs self-coupling via the renormalization group equation is considered, $\lambda(\phi)=\lambda_0+b_0\ln^2(\phi/\mu)$, 
it was shown that the results keep to be almost the same \cite{Yi:2020kmq}.
Therefore, the mechanism is not sensitive to the choices of model parameters.

\begin{table*}[htbp]
	\begin{tabular}{lcclllllll}
		\hline
		\hline
		Model \quad   &$h$& $w$ &$\phi_p/10^{-2}$ &$\phi_*/10^{-2}$&$f_0$ &$N$&$n_s$&$r$&$k_{\text{peak}}/\text{Mpc}^{-1}$\\
		\hline
 		H11w \quad   &$ 1.56\times10^{16}$~& $ 1.12\times 10^{-14}$ ~& $1.31$~&$1.38$~&$1.42\times 10^{48}$~&$56$~&$0.968$~&$0.038$~&$2.66\times 10^{12}$\\
        H12w \quad   &$ 2.713\times10^{17}$& $ 3.61\times 10^{-14}$ & $1.284$&$1.4$&$9.60\times 10^{47}$&$66$&$0.965$&$0.041$&$2.26\times 10^{12}$\\
        H11e \quad   &$ 1.415\times10^{16}$& $ 1.18\times 10^{-14}$ & $1.356$&$1.4$&$9.46\times 10^{47}$&$53$&$0.967$&$0.040$&$1.56\times 10^{9}$\\
        H12e \quad   &$ 7.77\times10^{16}$& $ 7.70\times 10^{-14}$ & $1.325$&$1.38$&$1.42\times 10^{48}$&$61$&$0.967$&$0.039$&$ 1.71\times 10^{8}$\\
        H11s \quad   &$ 1.47\times10^{15}$& $ 1.128\times 10^{-13}$ & $1.381$&$1.4$&$9.11\times 10^{47}$ &$52$&$0.970$&$0.041$&$2.86\times 10^{5}$\\
        H12s \quad   &$ 1.79\times10^{17}$& $ 3.51\times 10^{-14}$ & $1.354$&$1.38$&$1.42\times 10^{48}$ &$60$&$0.967$&$0.039$&$ 2.56\times 10^{5}$\\ \hline
        H11wf \quad  &$ 1.56\times10^{16}$~& $ 1.12\times 10^{-14}$ ~& $1.31$~&$1.38$~&$1.42\times 10^{48}$~&$56$~&$0.968$~&$0.038$~&$2.55\times 10^{12}$\\ \hline
        H21w \quad   &$ 9.147\times10^{13}$& $ 1.841\times 10^{-12}$ & $1.47$&$1.61$&$5.51\times 10^{24}$&$59$&$0.964$&$0.075$&$ 4.05\times 10^{12}$\\
        H22w \quad   &$ 2.526\times10^{17}$& $ 3.709\times 10^{-14}$ & $1.416$&$1.62$&$5.29\times 10^{24}$&$68$&$0.964$&$0.074$&$ 4.50\times 10^{12}$\\
        H21e \quad   &$ 8.70\times10^{13}$& $ 1.81\times 10^{-12}$&$1.535$& $1.61$&$5.69\times 10^{24}$&$59$&$0.967$&$0.072$&$ 1.18\times 10^{9}$\\
        H22e \quad   &$ 2.705\times10^{16}$& $ 1.83\times 10^{-13}$ & $1.51$&$1.62$&$5.28\times 10^{24}$&$63$&$0.964$&$0.074$&$9.57\times 10^{8} $\\
        H21s \quad   &$ 8.70\times10^{13}$& $ 1.78\times 10^{-12}$ & $1.578$&$1.61$&$5.51\times 10^{24}$ &$58$&$0.972$&$0.072$&$ 2.55\times 10^{5}$\\
        H22s \quad   &$ 2.70\times10^{16}$& $ 1.657\times 10^{-13}$ & $1.562$&$1.62$&$5.28\times10^{24}$&$62$&$0.964$&$0.073$&$3.56\times 10^{5}$\\
		\hline
		\hline
	\end{tabular}
	\caption{The chosen parameter sets and the predictions of $n_s$ and $r$ for the Higgs model. }
\label{higgst1}
\end{table*}

\begin{figure}[htbp]
  \centering
  \includegraphics[width=0.95\columnwidth]{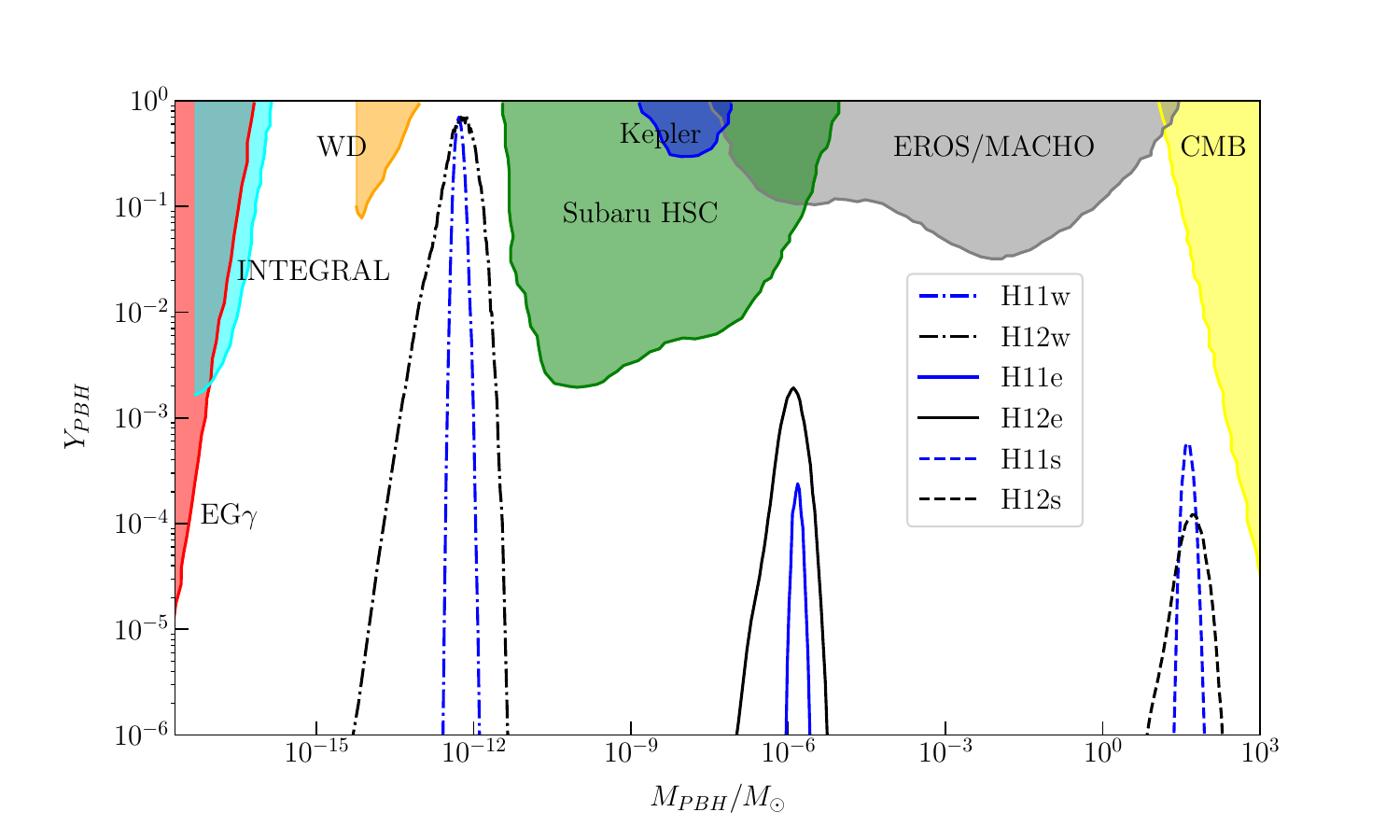}
  \caption{The abundances of PBHs produced in the models H11 and H12.
  The blue and black lines denote the results for the models H11 and H12, respectively. The shaded regions show the observational constraints on the PBH abundance:
the yellow region from accretion constraints by CMB \cite{Ali-Haimoud:2016mbv,Poulin:2017bwe},
the red region from extragalactic gamma-rays by PBH evaporation (EG$\gamma$) \cite{Carr:2009jm}, the cyan region from galactic center 511 keV gamma-ray line (INTEGRAL) \cite{Laha:2019ssq,Dasgupta:2019cae,Laha:2020ivk}, the orange region from white dwarf explosion (WD) \cite{Graham:2015apa},
the green region from microlensing events with Subaru HSC \cite{Niikura:2017zjd},
the blue region from the Kepler satellite \cite{Griest:2013esa},
the gray region from the EROS/MACHO \cite{Tisserand:2006zx}.}\label{higgs:fpbh}
\end{figure}

When the primordial curvature perturbation reenters the horizon during radiation dominated epoch,
if the energy density   contrast is large enough, it may gravitationally collapse to form PBHs.
The parameter to  describe  the abundance of PBHs is  the  current fractional energy density of PBHs with the mass $M$ to DM which is derived  in   Appendix \ref{app:pbh} in detail,
and it is \cite{Carr:2016drx,Gong:2017qlj}
\begin{equation}
\label{fpbheq1}
\begin{split}
Y_{\text{PBH}}(M)= &\frac{\beta(M)}{3.94\times10^{-9}}\left(\frac{\gamma}{0.2}\right)^{1/2}
\left(\frac{g_*}{10.75}\right)^{-1/4} \\
&\times \left(\frac{0.12}{\Omega_{\text{DM}}h^2}\right)
\left(\frac{M}{M_\odot}\right)^{-1/2},
\end{split}
\end{equation}
where $M_{\odot}$ is the solar mass, $\gamma= 0.2$ \cite{Carr:1975qj},  
$g_*$ is the effective degrees of freedom at the formation time, $\Omega_{\text{DM}}$ is the current energy density parameter of DM and $\Omega_{\text{DM}}h^2=0.12$ \cite{Aghanim:2018eyx};
the fractional energy density of PBHs at the formation related to the power spectrum is \cite{Young:2014ana, Ozsoy:2018flq,Tada:2019amh}
\begin{equation}
\label{pbh:beta}
  \beta(M) \approx \sqrt{\frac{2}{\pi}}\frac{\sqrt{\mathcal{P}_{\zeta}}}{\mu_c}
\exp\left(-\frac{\mu_c^2}{2\mathcal{P}_{\zeta}}\right),
\end{equation}
with $\mu_c=9\delta_c/2\sqrt{2}$ and the threshold of the  density perturbation for the PBH formation $\delta_c=0.4$ \cite{Musco:2012au,Harada:2013epa,Tada:2019amh,Escriva:2019phb,Yoo:2020lmg}. 
The value of $\delta_c$ may become larger by a factor of two if the non linearities between the Gaussian
curvature perturbation and the density contrast and the nonlinear effects arising at horizon crossing are considered \cite{Musco:2018rwt,Musco:2020jjb}. 
Furthermore, the abundance of PBHs depends on the shape and non-Gaussianity of $\mathcal{P}_{\zeta}$ and nonlinear statistics need to be taken into account \cite{Atal:2018neu,Germani:2018jgr,Germani:2019zez}.
It was shown in Ref. \cite{Zhang:2020uek} that non-Gaussianities of Higgs fluctuations are small at both the CMB and peak scales, 
so the effect of non-Gaussianities in the model is negligible.
Combining Eq. \eqref{fpbheq1}  and Eq. \eqref{pbh:beta}, and using the numerical results of the power spectra for the models in Table \ref{higgst1}
and Fig. \ref{pr:higgs}, we obtain the abundance of PBHs for the corresponding models,
and the results are shown in Fig. \ref{higgs:fpbh}. The mass scale and the peak abundance of PBHs are shown in Table \ref{higgst2}.

From Figs. \ref{pr:higgs} and \ref{higgs:fpbh}, we see that different peak scales correspond
to different peak masses of PBHs.
A broad peak in the primordial scalar power spectrum produces a broad peak in $Y_\text{PBH}$.
The models with the label ``w" produce PBHs with the mass
around  $10^{-14}-10^{-12} M_{\odot}$, the models with the label ``e" produce  PBHs with the Earth's mass, the models with the label ``s" produce  PBHs with the stellar mass.
The PBHs with masses around $10^{-14}-10^{-12} M_{\odot}$  almost   make up all the DM 
and the peak abundances of them are about $Y_\text{PBH}^\text{peak}\approx 1$.
Therefore, the Higgs field not only drives inflation but also explains DM.
In our models H11e, H12e, H21e and H22e, we successfully produce PBHs with the mass around $\mathcal{O}(1)M_{\oplus}$ which could be used to explain the origin of the Planet 9.
PBHs with the stellar mass which explain the LIGO events are also produced in the Higgs model.

\begin{table}[htp]
	\renewcommand\tabcolsep{5.0pt}
	\begin{tabular}{lllllllllll}
		\hline
		\hline
		Model \quad   &$\mathcal{P}_{\zeta(\text{peak})}$& $M_\text{peak}/M_\odot$&$Y_\text{PBH}^\text{peak}$& $f_c/\text{Hz}$\\
		\hline
 		H11w \quad   &$0.0128$&$5.22\times10^{-13}$&$0.70$&$4.30\times 10^{-3}$\\
        H12w \quad   &$0.0129$&$7.22\times10^{-13}$&$0.73$&$ 4.01\times 10^{-3}$\\
        H11e \quad   &$0.0126$&$1.51\times 10^{-6}$&$2.38\times10^{-4}$&$2.70\times10^{-6}$\\
        H12e \quad   &$0.0134$&$1.26\times 10^{-6}$&$1.92\times10^{-3}$&$ 2.90\times 10^{-6} $\\
        H11s \quad   &$0.0178$&$45.1 $&$5.72\times10^{-4}$&$4.82\times10^{-10}$\\
        H12s \quad   &$0.0167$&$50.8 $&$1.22\times10^{-4}$&$4.37\times10^{-10}$\\
        \hline
       	H11wf \quad   &$0.0129$&$5.65\times10^{-13}$&$0.78$&$4.61\times 10^{-3}$\\
        \hline
        H21w \quad   &$0.0128$&$2.24\times10^{-13}$&$0.97$&$6.64\times 10^{-3}$\\
        H22w \quad   &$0.0126$&$1.82\times10^{-13}$&$0.63$&$6.96 \times 10^{-3}$\\
        H21e \quad   &$0.0131$&$2.66\times 10^{-6}$&$6.37\times10^{-4}$&$2.03\times10^{-6}$\\
        H22e \quad   &$0.0120$&$4.02\times 10^{-6}$&$2.55\times10^{-5}$&$1.48\times10^{-6}$\\
        H21s \quad   &$0.0176$&$56.5 $&$3.97\times10^{-4}$&$6.69\times10^{-10}$\\
        H22s \quad   &$0.0173$&$21.9 $&$4.01\times10^{-4}$&$6.33\times10^{-10}$\\
		\hline
		\hline
	\end{tabular}
	\caption{The results for the peak amplitude of the primordial scalar power spectrum, the peak mass and   abundance of PBHs and the peak frequency of SIGWs with the chosen parameter sets shown in Table \ref{higgst1}.}
\label{higgst2}
\end{table}

\begin{figure}[htbp]
  \centering
  \includegraphics[width=0.95\columnwidth]{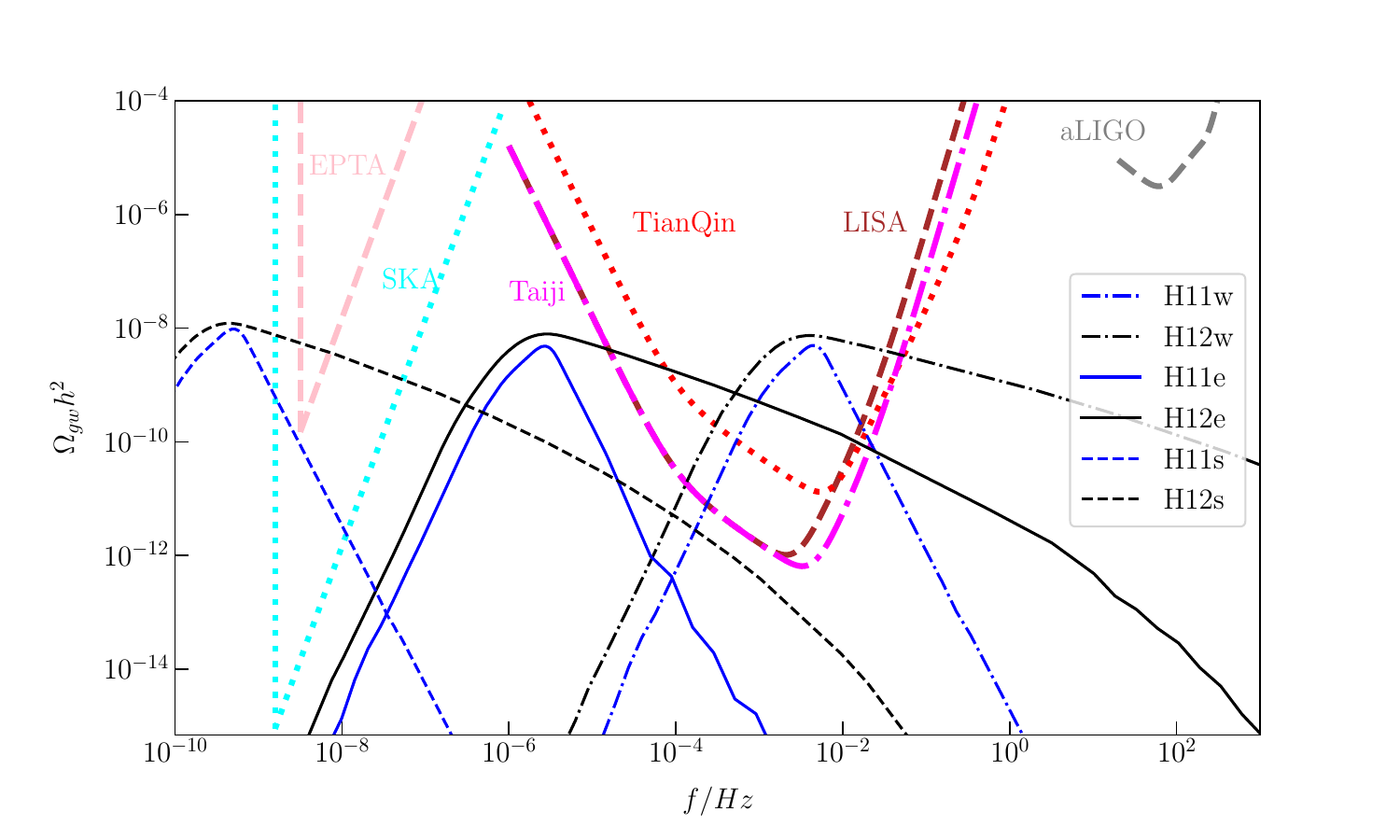}
  \caption{The scalar induced secondary GWs for the Higgs model. The blue and  black   lines denote the models H11 and H12, respectively.
  The pink dashed curve denotes the EPTA limit \cite{Ferdman:2010xq,Hobbs:2009yy,McLaughlin:2013ira,Hobbs:2013aka} ,
the cyan dotted curve denotes the SKA limit \cite{Moore:2014lga},
the red dot-dashed curve in the middle denotes the TianQin limit \cite{Luo:2015ght},
the dotted magenta curve shows the Taiji limit \cite{Hu:2017mde},
the brown dashed curve shows the LISA limit \cite{Audley:2017drz},
and the gray dot-dashed curve denotes the aLIGO limit \cite{Harry:2010zz,TheLIGOScientific:2014jea}.}\label{gwhiggs}
\end{figure}

Accompanied by the production of PBHs, the  large scalar perturbations induce GWs  during radiation.
These SIGWs consist of the stochastic background,
and they  could be detected by the space based GW detectors like LISA \cite{Danzmann:1997hm,Audley:2017drz}, Taiji \cite{Hu:2017mde} and TianQin  \cite{Luo:2015ght}
in the future and provide  additional constraints on the primordial Universe. 
We present the  energy density of SIGWs  in the radiation domination in detail in the Appendix  \ref{app:sgw}, and it is  \cite{Inomata:2016rbd,Kohri:2018awv}
\begin{equation}
\label{gwres1}
\begin{split}
\Omega_{\mathrm{GW}}(k,\eta)=&\frac{1}{6}\left(\frac{k}{aH}\right)^2\int_{0}^{\infty}dv
\int_{|1-v|}^{1+v}du\\ 
&\left[\frac{4v^2-(1-u^2+v^2)^2}{4uv}\right]^2\\
&\times \overline{I_{\text{RD}}^{2}(u, v, x)} \mathcal{P}_{\zeta}(kv)\mathcal{P}_{\zeta}(ku),
\end{split}
\end{equation}
where $I_\text{RD}$ is defined in Eq. \eqref{irdeq1}.
Substituting the numerical results for the power spectra of the models listed in Table \ref{higgst1} into  Eq. \eqref{gwres1},
we obtain the energy density $\Omega_{\text{GW}}$  of SIGWs  for the corresponding models and the results are shown in Fig. \ref{gwhiggs}.
The peak frequencies $f_c$ of these SIGWs are shown in Table \ref{higgst2}.
From Table \ref{higgst2}, we see that the peak frequencies of the SIGWs are around mHz, $10^{-6}$Hz and nHz respectively.
For the models H11 and H21, both $\mathcal{P}_\zeta$ and $\Omega_{\text{GW}}$ have sharp peaks.
The mHz and nHz SIGWs may be detected by LISA/Taiji/TianQin and PTA respectively.
For the models H12 and H22, the enhanced $\mathcal{P}_\zeta$ has half-domed shape and $\Omega_{\text{GW}}$ has a broad shape which spans a wide frequency bands.
The models H12s and H22s which produce the stellar mass PBHs and have a broad shape for $\Omega_{\text{GW}}$ are excluded by the EPTA data \cite{Ferdman:2010xq,Hobbs:2009yy,McLaughlin:2013ira,Hobbs:2013aka},
while the models H11s and H21s generating SIGWs with a sharp peak can be tested by SKA \cite{Moore:2014lga}.
The models H11w, H12w, H12e, H21w, H22w and H22e can be tested by LISA/Taiji/TianQin.
However, the models H11e and H21e which generate SIGWs with a sharp peak is beyond the reach of either SKA or LISA/Taiji/TianQin.

From the point of view of the effective field theory,
the lower dimensional terms in $f(\phi)$ need to be taken into account,
i.e., we need to consider $f(\phi)=f_0\phi^{22}+\delta f$ with
$\delta f=\sum_{n=4}^{21}C_n \phi^n$ and $C_n<O(1)$.
To consider the effects of the lower dimensional terms of $f(\phi)$,
we choose $\delta f=\sum_{n=4}^{21}\phi^n$ and as an example we apply the correction $\delta f$ to the model H11w.
The model with the correction $\delta f$ in $f(\phi)$ is called H11wf. 
We choose the same parameters for the models H11w and H11wf and the results are shown in Tables \ref{higgst1} and \ref{higgst2}.
From Tables \ref{higgst1} and \ref{higgst2}, 
we see that the results for the models H11wf and H11w are almost the same.
Therefore, the lower dimensional terms have little effect on the enhancement mechanism
and the correction $\delta f$ does not spoil the enhancement mechanism.

\section{T-model}
T-model with the canonical kinetic term is consistent with the observational constraints.
In this section, we show that our mechanism works for T-model too,
henceforth show that the inflationary potential in our mechanism is not restricted.
The potential for T-model is \cite{Kallosh:2013maa,Kallosh:2013hoa,Yi:2016jqr}
\begin{equation}\label{T:model}
V(\phi)=V_0\tanh^{2m}\left(\frac{\phi}{\sqrt{6\alpha}}\right).
\end{equation}
Substituting the potential \eqref{T:model}  into Eq. \eqref{gfuncn}, we obtain
\begin{equation}\label{tmodel:fn}
  f(\phi)=f_0~ \text{sech}^4\left(\frac{\phi}{\sqrt{6\alpha}}\right) \tanh^{-2+4m/n}\left(\frac{\phi}{\sqrt{6\alpha}}\right),
\end{equation}
where
\begin{equation}\label{tmodel:fn0}
f_0=\frac{2}{3\alpha} \left(\frac{V_0}{U_0}\right)^{2/n}\frac{m^2}{n^2}.
\end{equation}
For the parameters considered in this section, we take $f_0= 36$. For  $n=1/3$,  we obtain
\begin{equation}\label{tmodel:f1}
 f(\phi)= 36 \text{sech}^4\left(\frac{\phi}{\sqrt{6\alpha}}\right)\tanh^{12m-2}\left(\frac{\phi}{\sqrt{6\alpha}}\right).
\end{equation}
For  $n=2/3$,  we obtain
\begin{equation}\label{tmodel:f2}
f(\phi)=36 \text{sech}^4\left(\frac{\phi}{\sqrt{6\alpha}}\right)\tanh^{6m-2}\left(\frac{\phi}{\sqrt{6\alpha}}\right).
\end{equation}
Similar to the Higgs model, at low energy scales $\phi\ll1$, $f(\phi)$ is negligible and the model reduces to the standard case
with the canonical kinetic term.

As discussed above, the peak shape of the enhanced power spectrum is determined by $q$ in the peak function \eqref{gfuncng}. Similar to the Higgs model, we also consider the peak function \eqref{gfuncng} with $q = 1$ and $q = 5/4$ to explore the sharp and broad peaks. 
To distinguish different models, 
as shown in Table \ref{tmodellabel}, 
the model with $m=1/6$, $\alpha=1$ and $n=1/3$ is labeled as T1
and the model with $m=1/3$, $\alpha=1$ and $n=2/3$ is labeled as T2. 
Additionally, we label the model with $n = 1/3$
and $q=1$ as T11, the model with $n=1/3$ and  $q=5/4$ as T12, the model with $ n=2/3$
and $q=1$ as T21, and the model with $n =2/3$ and  $q =5/4$ as T22, respectively.
\begin{table}[htbp]
	\begin{tabular}{llll}
		\hline
		\hline
		Label \quad  & $U(\phi)=U_0\Phi^n$ \quad\quad&  $q$ \\
		\hline
 		T11 \quad    & $n=1/3$  & $q=1$\\
 		T12 \quad    & $n=1/3$  & $q=5/4$\\
 		T21 \quad    & $n=2/3$   &$q=1$\\
 		T22 \quad    &$n=2/3$  & $q=5/4$\\
		\hline
		\hline
	\end{tabular}
	\caption{The labeling for the T-models with different peak functions and effective potentials.}
\label{tmodellabel}
\end{table}

\begin{figure}[htbp]
  \centering
  \includegraphics[width=0.95\columnwidth]{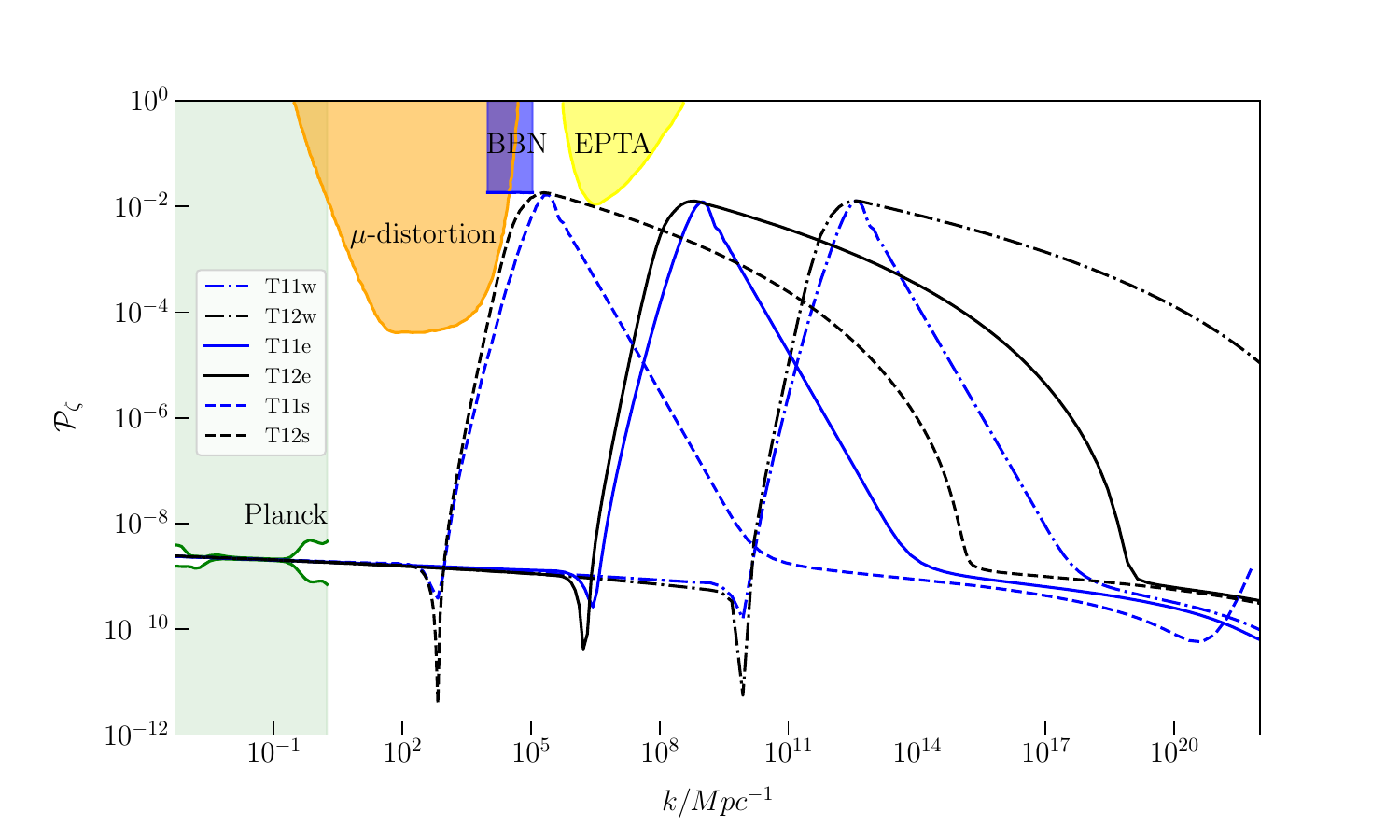}
  \caption{The enhanced power spectrum in the T-model. The blue and black lines
  denote the results for the models T11 with sharp peak and T12 with broad peak, respectively.}\
  \label{pr:tmodel}
\end{figure}

Choosing the parameter $V_0$, the value $\phi_*$ of the scalar field  at the pivotal scale, and  the parameters $h$, $w$ and $\phi_p$ in the peak function as shown in Table
\ref{tmodel:t1},
we numerically solve the background equations \eqref{Eq:eom1}-\eqref{Eq:eom3} and the perturbation equation \eqref{zeta:k} to obtain the scalar power spectrum and the results are shown in Fig. \ref{pr:tmodel}.
The power spectra for the models T11 and T12 are denoted by blue and black lines respectively.
The results for the models T21 and T22 are similar with those of T11 and T12,
so we do not show them in the figures.
The power spectrum
produced in the model T11 has a sharp peak while the power spectrum produced in the
model T12 has a broad peak.  This implies that the peak shape in the power spectrum is unaffected by the potential but affected by the peak function.
The scale where the power spectrum is enhanced  can be adjusted by  the parameter $\phi_p$  and the results are shown in Table \ref{tmodel:t1} and Fig \ref{pr:tmodel}.
The numerical results for the  scalar spectral index $n_s$ and the tensor-to-scalar ratio
$r$ are shown in Table \ref{tmodel:t1}.
For the models T11 and T12, we get
\begin{equation}
    n_s\approx 0.968, \quad r\approx 0.038.
\end{equation}
These results are similar to those for the Higgs models H11 and H12.
As expected,
$n_s$ and $r$ are almost unaffected by the peak function whose effect is negligible at large scales
as well as the  inflationary potential, and they are determined by the effective potential.

\begin{table*}[htbp]
	\begin{tabular}{lccccccccll}
		\hline
		\hline
		Model \quad   &$h$& $w$ &$\phi_p$ &$\phi_*$&$U_0/10^{-9}$&  $N$&$n_s$&$r$&$k_{\text{peak}}/\text{Mpc}^{-1}$\\
		\hline
 		T11w \quad   &$ 4.86\times10^{9}$~& $ 8.885\times 10^{-11}$ ~& $0.44$~&$0.81$~&$1.67$ ~&$56$~&$0.969$~&$0.037$~&$4.01\times 10^{12}$\\
        T12w \quad   &$ 2.109\times10^{11}$& $ 8.481\times 10^{-11}$ & $0.32$&$0.81$&$1.71$ &$66$&$0.966$&$0.038$&$3.85\times 10^{12}$\\
        T11e \quad   &$ 3.57\times10^{9}$& $ 8.89\times 10^{-11}$ & $0.58$&$0.81$&$1.65$ &$55$&$0.969$&$0.037$&$1.01\times 10^{9}$\\
        T12e \quad   &$ 1.139\times10^{9}$& $ 8.68\times 10^{-11}$ & $0.51$&$0.81$&$1.71 $& $61$&$0.967$&$0.038$&$6.12\times 10^{8}$\\
         T11s \quad   &$ 3.178\times10^{9}$& $ 8.92\times 10^{-11}$ & $0.644$&$0.76$&$1.84$ &$51$&$0.970$&$0.040$&$2.31\times 10^{5} $\\
        T12s \quad   &$ 8.633\times10^{9}$& $ 8.56\times 10^{-11}$ & $0.642$&$0.81$&$1.70$ &$60$&$0.967$&$0.038$&$2.00\times 10^{5}$\\
        \hline
        T21w \quad   &$ 9.205\times10^{9}$& $ 4.85\times 10^{-11}$ & $0.80$&$1.31$&$3.07$ &$61$&$0.969$&$0.063$&$3.02\times 10^{12}$\\
        T22w \quad   &$ 4.51\times10^{11}$& $ 4.644\times 10^{-11}$ & $0.63$&$1.30$&$3.23$ &$70$&$0.967$&$0.066$&$5.47\times 10^{12}$\\
        T21e \quad   &$ 8.34\times10^{9}$& $ 4.17\times 10^{-11}$ & $0.98$&$1.31$&$3.04$ &$61$&$0.970$&$0.062$&$8.20\times 10^{8}$\\
        T22e \quad   &$ 2.33\times10^{11}$& $ 5.54\times 10^{-11}$ & $0.88$&$1.31$&$3.17$ &$66$&$0.967$&$0.065$&$7.97\times 10^{8}$\\
        T21s \quad   &$ 7.00\times10^{9}$& $ 4.11\times 10^{-11}$ & $1.14$&$1.31$&$2.97$ &$60$&$0.974$&$0.061$&$2.73\times 10^{5}$\\
        T22s \quad   &$ 2.20\times10^{11}$& $ 4.74\times 10^{-11}$ & $1.06$&$1.31$&$3.16$ &$65$&$0.967$&$0.065$&$2.31\times 10^{5}$\\
		\hline
		\hline
	\end{tabular}
	\caption{The chosen parameter sets and the predictions of $n_s$ and $r$ for the T-model.}
\label{tmodel:t1}
\end{table*}

\begin{figure}[htbp]
  \centering
  \includegraphics[width=0.95\columnwidth]{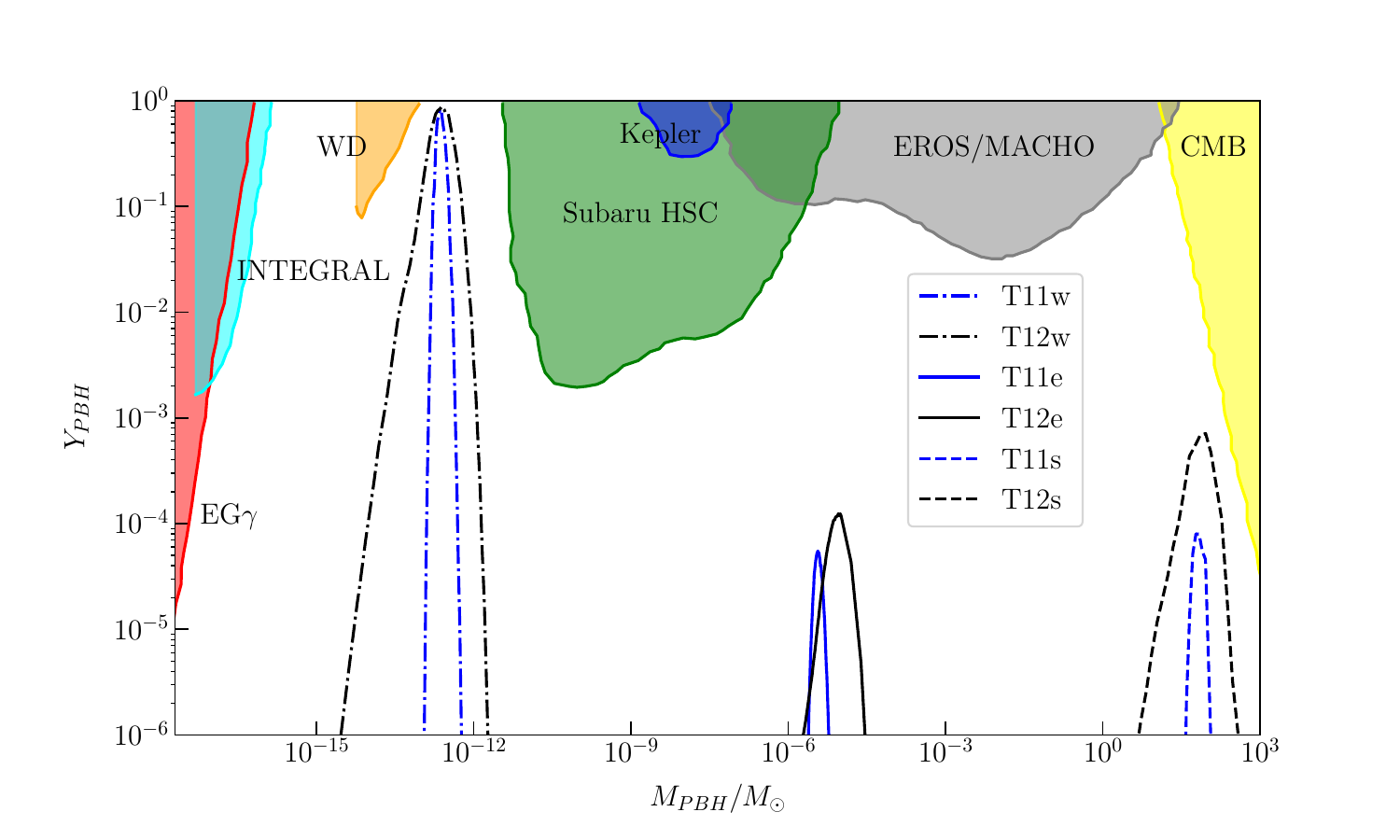}
  \caption{The PBHs abundance for the models display in Fig. \ref{pr:tmodel}. The blue   and black   lines    denote the models labeled as T11, T12, respectively.  }\label{fpbh:tmodel}
\end{figure}

Using  the numerical results of the power spectra for the models as shown in Fig. \ref{pr:tmodel} and combining them with  Eq. \eqref{fpbheq1},
we obtain the  abundances of   PBHs and the results are  shown in  Fig. \ref{fpbh:tmodel}.
The peak mass and the abundance of PBHs are shown in Table \ref{tmodelt2}.
The models labeled as ``w" produce PBHs with the mass around $10^{-14}-10^{-12} M_{\odot}$ and the  peak  abundance  $Y_\text{PBH}^\text{peak}\approx 1$,
which  make up almost all the DM.
The models labeled as ``e" produce PBHs with the mass around $\mathcal{O}(1)M_{\oplus}$, which can be used to  explain the origin of the Planet 9.
The models labeled as ``s" produce PBHs with the stellar mass that may account for the LIGO events.
Therefore, in our mechanism,
T-model can also produce PBHs with different mass and explain DM in terms of PBHs.

\begin{table}[htp]
	\renewcommand\tabcolsep{4.0pt}
	\begin{tabular}{lllllllllll}
		\hline
		\hline
		Model \quad   &$\mathcal{P}_{\zeta(\text{peak})}$& $M_\text{peak}/M_\odot$&$Y_\text{PBH}^\text{peak}$& $f_c/\text{Hz}$\\
		\hline
 		T11w \quad   &$0.0127$&$2.28\times10^{-13}$&$0.83$&$7.03\times 10^{-3}$\\
        T12w \quad   &$0.0128$&$2.48\times10^{-13}$&$0.88$& $6.54\times 10^{-3} $\\
        T11e \quad   &$0.0122$&$ 3.67\times 10^{-6}$&$5.51\times10^{-5}$&$1.70\times 10^{-6}$\\
        T12e \quad   &$0.0127$&$9.04\times 10^{-6}$&$1.24\times10^{-4}$&$1.07\times 10^{-6}$\\
        T11s \quad   &$0.0165$&$59.97 $&$8.00\times10^{-5}$&$3.83\times 10^{-10}$\\
        T12s \quad   &$0.0183$&$91.75 $&$7.12\times10^{-4}$&$3.48\times 10^{-10}$\\
        \hline
        T21w \quad   &$0.0128$&$4.03\times10^{-13} $&$0.77$&$5.06\times 10^{-3}$\\
        T22w \quad   &$0.0125$&$ 9.71\times10^{-14} $&$0.58$&$9.91\times10^{-3}$\\
        T21e \quad   &$ 0.0127 $&$5.47\times10^{-6}$&$1.35\times 10^{-4}$&$1.43\times 10^{-6}$\\
        T22e \quad   &$0.0126$&$5.79\times10^{-6} $&$1.21\times 10^{-4}$&$3.21\times 10^{-6}$\\
        T21s \quad   &$0.0166$&$49.4 $&$9.73\times10^{-5}$&$5.10\times 10^{-10}$\\
        T22s \quad   &$0.0174$&$45.4 $&$3.10\times 10^{-4}$&$ 4.40\times 10^{-10} $\\
		\hline
		\hline
	\end{tabular}
	\caption{The results for the peak amplitude of the primordial scalar power spectrum, the peak mass and   abundance of PBH and the peak frequency of SIGWs with the chosen parameter sets shown in Table \ref{tmodel:t1}.}
\label{tmodelt2}
\end{table}

\begin{figure}[htbp]
  \centering
  \includegraphics[width=0.95\columnwidth]{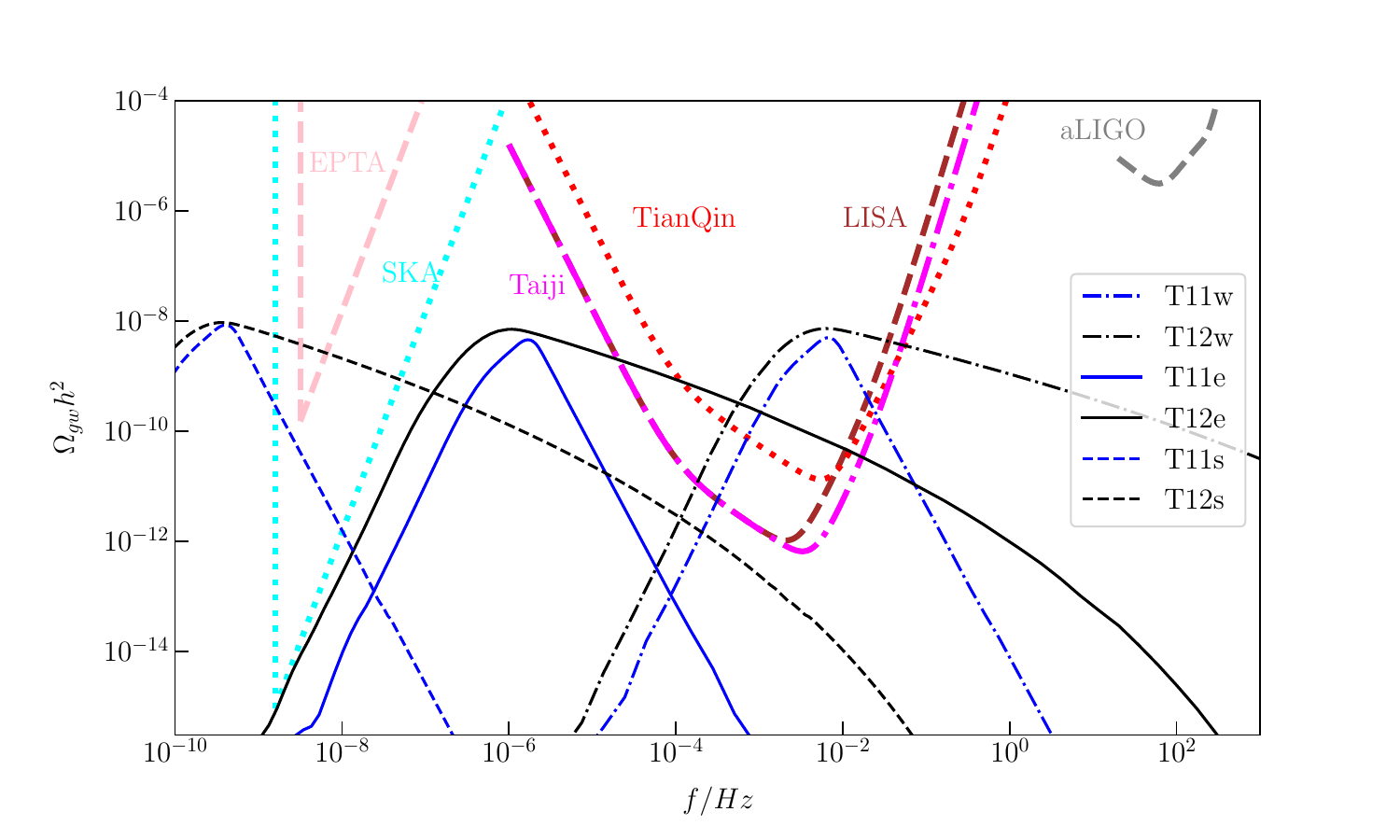}
  \caption{The SIGWs for the T-model.  The blue and black lines denote the models T11 and T12, respectively.}
  \label{gw:tmodel}
\end{figure}

Combining the numerical results of the power spectra as shown in Fig. \ref{pr:tmodel} and Eq. \eqref{gwres1},
we obtain the energy density $\Omega_\text{GW}$ of SIGWs and the results are shown in Fig. \ref{gw:tmodel}.
The results are similar to those in the Higgs model.
The peak frequencies of these SIGWs are shown in Table \ref{tmodelt2}.
The SIGWs  have a sharp peak in the models T11 and have a broad peak in the models T12.
The scalar perturbations producing PBHs with the mass around $10^{-14}-10^{-12} M_{\odot}$ generate SIGWs with the frequency around mHz which can be detected by LISA/Taiji/TianQin.
The scalar perturbations producing PBHs with the mass around $M_{\oplus}$ generate SIGWs with the frequency around $10^{-6}$Hz.
The SIGWs with a broad peak may be  detected by LISA/Taiji/TianQin,
but the SIGWs with  a sharp peak is beyond the reach of either SKA  or LISA/Taiji/TianQin.
The scalar perturbations producing PBHs with the stellar mass generate SIGWs with the frequency around nHz.
The SIGWs with a broad peak are excluded by the EPTA data
and the SIGWs with a sharp peak can be tested by SKA.

\section{conclusion}

In the standard case that the inflaton has the canonical kinetic term and is
minimally coupled to gravity, Higgs field as the inflaton is excluded by
the CMB constraints because the amplitude of the primordial GWs produced in
Higgs model is too large.
By introducing a noncanonical kinetic term with a peak, the amplitude of
the primordial GWs produced in Higgs model at large scales is successfully reduced and the inflation driven by the Higgs field is consistent with Planck 2018 results.
The scalar power spectrum is also enhanced at small scales.
The enhanced curvature perturbations at small scales produce abundant PBH DM and generate SIGWs after the horizon reentry during radiation domination.
The peak scale at which the scalar power spectrum is enhanced can be easily adjusted.
As examples, we use three different parameter sets to show how to get the enhanced power spectra at three different scales. With these enhanced power spectra,
PBHs with the peak mass around $10^{-14}-10^{-12} M_{\odot}$, the Earth's mass and the stellar mass are produced respectively.
The corresponding peak abundances are $Y_{\text{PBH}}\sim 1$, $Y_{\text{PBH}}\sim 10^{-4}$
and $Y_{\text{PBH}}\sim 10^{-4}$. Therefore, PBHs with the peak mass around $10^{-14}-10^{-12} M_{\odot}$ can explain almost all the DM.
At the same time, the SIGWs with the peak frequency
around mHz, $10^{-6}$Hz and nHz are generated. Dependent on the peak function,
the peak shape of the enhanced power spectrum can be sharp or broad which is like half domed,
and the peak shape of the energy density of the SIGWs can be also sharp or broad.
The SIGWs with broad peaks and peak frequency around nHz are excluded by the EPTA observations.
In general, the SIGWs can be tested by PTA observations and space based GW detectors.

The scalar spectral index $n_s$ and  the tensor-to-scalar ratio $r$ predicated in the T-model with the canonical kinetic term are consistent with Planck 2018 results.
We also show that the enhancement mechanism works for the T-model and
the results for the T-model are similar to those for the Higgs model.

In conclusion, with the enhancement mechanism provided by the noncanonical kinetic term with a peak,
the Higgs field accounts for the origin of mass and DM in terms of PBHs,
the SIGWs can be tested by PTA observations and the space based GW detectors.
In addition to the Higgs field, the enhancement mechanism also works for T-model and other inflationary potentials.
The energy density of the SIGWs can have sharp or broad peaks.
Both the GW and PBH observations can be used to test this mechanism and probe physics in the early universe.

\begin{acknowledgments}
This research is supported in part by the National Natural Science
Foundation of China under Grant No. 11875136,
the Major Program of the National Natural Science Foundation of China under Grant No. 11690021 and the National Key Research and Development Program of China under Grant No. 2020YFC2201504. 
Z. Y. is supported by China Postdoctoral Science Foundation Funded Project under Grant No. 2019M660514. 
Z. H. Z. is supported by the National Natural Science Foundation of China under Grants No. 11633001, No. 11920101003 and No. 12021003, the Strategic Priority Research Program of the Chinese Academy of Sciences, Grant No. XDB23000000 and the Interdiscipline Research Funds of Beijing Normal University.
\end{acknowledgments}

\appendix
\section{THE PRIMORDIAL BLACK HOLES}
\label{app:pbh}

The stellar mass PBHs could be those detected by LIGO/Virgo collaboration
and they can also explain DM,  even make up all the DM with the mass in the windows $10^{-17}-10^{-15}M_{\odot}$ and  $10^{-14}-10^{-12}M_{\odot}$.
In this appendix, we briefly review  the formulas about the production of PBHs.
The energy density fraction of PBHs at formation to the total mass of the Universe  is
\begin{equation}\label{beta}
  \beta=\left.\frac{\rho_{\text{PBH}}}{\rho }\right|_{\text{at formation}},
\end{equation}
where $\rho$ is the energy density of the Universe and  $\rho_{\text{PBH}}$ is the energy density of PBHs.
The relation between the energy density of PBHs at the formation and that at present is
\begin{equation}\label{pbh:rho}
  \rho_\text{PBH}=\left(\frac{a_0}{a}\right)^3 \frac{3H_0^2}{8\pi G}~\Omega_{\text{PBH}_0},
\end{equation}
where the  subscript `$0$' denotes  the present value  of a quantity and $\Omega_{\text{PBH}}=8\pi G \rho_\text{PBH}/ 3H^2 $ is the density parameter of PBHs. The relation between the energy density of the radiation  $\rho_r $ and temperature $ T$ is
\begin{equation}\label{rho:T}
  \rho_r =\frac{\pi^2}{30}g_* T^4,
\end{equation}
where $g_*$ is the number of relativistic degrees of freedom.
Assuming   the entropy is  conserved, we have
\begin{equation}\label{entropy:T}
S=g_{*s}a^3T^3=\text{constant},
\end{equation}
where $g_{*s}$ is  the number of entropy degrees of freedom  which is  approximately equal to
the  number of relativistic degrees of freedom, $g_{*s}=g_{*}$. Combining the above equations with  the  Friedmann equation of the Universe,
\begin{equation}\label{fri:eq}
  H^2=\frac{8\pi G}{3} \rho,
\end{equation}
and assuming PBHs are formed during the radiation dominated epoch, we obtain the relation of Hubble parameter at formation to that at today
\begin{equation}\label{H:H0}
 H^2=a^{-4}\left(\frac{g_*}{g_{*_0}}\right)^{-1/3}H_0^2\Omega_{r_0},
\end{equation}
where  we set $a_0=1$. Combining Eq. \eqref{fri:eq} with Eq. \eqref{H:H0}, and using the definition \eqref{beta}, we obtain
\begin{equation}\label{beta2}
\beta=\left(\frac{H_0}{H}\right)^{1/2}\left(\frac{g_{*}}{g_{*_0}}\right)^{1/4}\Omega_{r_0}^{-3/4}\Omega_{\text{PBH}_0}.
\end{equation}
The mass of the PBH could be assumed as
\begin{equation}\label{pbhmass}
  M_{\text{PBH}}=\gamma M_H,
\end{equation}
where $\gamma\approx 0.2$ \cite{Carr:1975qj} is determined by the detail of the formation of the PBH and  $M_H$ is the total mass in the horizon at the PBH formation,
\begin{equation}\label{mass:h}
  M_H=\frac{4\pi}{3}H^{-3} \rho =\frac{1}{2GH}.
\end{equation}
At the peak scale $k_\text{peak}$, the peak mass of PBH is
\begin{equation}\label{mass:pm}
  M_\text{peak}=3.68\left(\frac{\gamma}{0.2}\right) \left(\frac{g_*}{10.75}\right)^{-1/6} \left(\frac{k_\text{peak}}{10^6\ \text{Mpc}^{-1}}\right)^{-2} M_\odot.
\end{equation}
Combining   Eq. \eqref{beta2} and Eq. \eqref{pbhmass}, we obtain
\begin{equation}\label{beta3}
  \beta=\gamma^{-1/2} \left(\frac{M_\text{PBH}}{M_0}\right)^{1/2}\left(\frac{g_{*}}{g_{*_0}}\right)^{1/4}\Omega_{r_0}^{-3/4}\Omega_{\text{PBH}_0},
\end{equation}
where  $M_0=(2GH_0)^{-1}$ is the horizon mass at the present.
The  definition for the   current fractional energy density of PBHs with the mass $M$ to the DM is
\begin{equation}\label{pbh:y}
  Y_\text{PBH}=\frac{\Omega_{\text{PBH}_0}}{\Omega_{{\text{DM}_0}}}.
\end{equation}
Substituting the definition \eqref{pbh:y} into \eqref{beta3}, we obtain
\begin{equation}\label{beta:y}
\beta=\gamma^{-1/2} \left(\frac{M_\text{PBH}}{M_0}\right)^{1/2}\left(\frac{g_{*}}{g_{*_0}}\right)^{1/4}\Omega_{r_0}^{-3/4}\Omega_{\text{DM}_0} Y_\text{PBH}.
\end{equation}

The PBHs would be formed if  the density contrast of the overdense regions at horizon reentry during radiation domination exceeds  a certain  threshold value. In Press-Schechter theory, the  fraction $\beta$  could be regarded as the probability that the density contrast exceeds  the threshold $\delta_c=0.4$ \cite{Harada:2013epa},
\begin{equation}\label{beta:pro}
  \beta=2\int_{\delta_c}^{1}P(\delta(R)) d\delta(R),
\end{equation}
where $\delta(R)$ is the smoothed density contrast and $R=(aH)^{-1}$ is the smooth scale,  $P(\delta(R))$ is the  distribution of the smoothed  overdensity. Assuming the Gaussian initial perturbations, the  probability function for the smoothed density contrast  is
\begin{equation}\label{gauss:pro}
  P(\delta(R))=\frac{1}{\sqrt{2\pi} \sigma(R)}\exp\left(-\frac{\delta(R)^2}{2\sigma(R)^2}\right),
\end{equation}
with   the  variance satisfying
\begin{equation}\label{simga1}
\sigma(R)^2=\int_{0}^{\infty}W^2(k R)\frac{\mathcal{P}_\delta(k)}{k}dk,
\end{equation}
where $\mathcal{P}_\delta$ is the power spectrum of the matter perturbation and $W(k R)$ is the window function. In our paper, we choose  the Gaussian  window function
\begin{equation}\label{Gauus:window}
 W(k R)=\exp\left(-k^2R^2/2\right),
 \end{equation}
other kinds of window functions are also possible \cite{Ando:2018qdb}. The relation between the matter perturbation and the primordial curvature perturbation is
\begin{equation}\label{rel:pp}
\mathcal{P}_\delta(k)=\frac{4(1+w)^2}{(5+3w)^2}\left(\frac{k}{aH}\right)^4 \mathcal{P}_{\zeta}(k).
\end{equation}
Substituting the Gaussian window function \eqref{Gauus:window} and Eq. \eqref{rel:pp} into Eq. \eqref{simga1}, the mass variance becomes
\begin{equation}\label{phb:variance}
\sigma^2=\frac{4(1+w)^2}{(5+3w)^2}\int_{0}^{\infty}x^3 \exp\left(-x^2\right) \mathcal{P}_{\zeta}(x/R)dx,
\end{equation}
where $x=k R$. As a suitable approximation, 
we assume the primordial curvature  power spectrum is scale invariant even though it may change rapidly, 
because the integral \eqref{phb:variance} is dominated by the scale $x\sim1$ and other scale has little influence on the result \cite{Sato-Polito:2019hws}. 
The mass variance \eqref{phb:variance} becomes
\begin{equation}\label{phb:variance2}
\sigma^2\approx\frac{2(1+w)^2}{(5+3w)^2} \mathcal{P}_{\zeta}(1/R).
\end{equation}
Substituting Eq. \eqref{phb:variance2} into the definition  \eqref{beta:pro} and combining it with the probability function  \eqref{gauss:pro}, we obtain
\begin{equation}
\label{pbh:beta2}
\begin{split}
\beta&=\text{erfc}\left(\frac{5+3w}{2(1+w)}\frac{\delta_c}{\sqrt{\mathcal{P}_\zeta}}\right) \\
&\approx \sqrt{\frac{2}{\pi}}\frac{\sqrt{\mathcal{P}_{\zeta}}}{\mu_c}
\exp\left(-\frac{\mu_c^2}{2\mathcal{P}_{\zeta}}\right),
\end{split}
\end{equation}
where $\text{erfc}(x)$ is the complementary error function, and
\begin{equation}
\mu_c=\frac{5+3w}{\sqrt{2}(1+w)} \delta_c.
\end{equation}
Combining Eq. \eqref{beta:y} and Eq. \eqref{pbh:beta2}, we can predict the abundance of PBHs produced from  the primordial curvature perturbations. Substituting $\Omega_{r_0}=9.17\times 10^{-5}$, $g_{*_0}=3.36$, $M_0\approx 4.63\times 10^{22}M_\odot$ and $H_0=67.27\text{km}/\text{s}/\text{Mpc}$ into Eq. \eqref{beta3}, we obtain
\begin{equation}
\label{pbh:fpbheq1}
\begin{split}
Y_{\text{PBH}}(M)= &\frac{\beta(M)}{3.94\times10^{-9}}\left(\frac{\gamma}{0.2}\right)^{1/2}
\left(\frac{g_*}{10.75}\right)^{-1/4} \\
&\times \left(\frac{0.12}{\Omega_{\text{DM}}h^2}\right)
\left(\frac{M}{M_\odot}\right)^{-1/2},
\end{split}
\end{equation}
where $M$ is the mass of the PBH.

\section{THE SCALAR INDUCED SECONDARY GRAVITATIONAL WAVES}
\label{app:sgw}
Accompanied by the production of PBHs,
the large scalar perturbations generate SIGWs during radiation domination.
In this appendix, we review  the  generation of SIGWs in detail.
The perturbed metric in the  Newtonian gauge in the cosmological background is
\begin{equation}
\begin{split}
    d s^2=&-a^2(\eta)(1+2\Phi)d\eta^2+\\
    & a^2(\eta)\left[(1-2\Phi)\delta_{ij}+\frac12h_{ij}\right]d x^i d x^j,
\end{split}
\end{equation}
where $\eta$ is the conformal time, $\Phi$ is the Bardeen potential, and we neglect the anisotropic
stress.  The Fourier transform of tensor perturbations $h_{ij}$ is
\begin{equation}
\label{hijkeq1}
h_{ij}(\bm{x},\eta)=\frac{1}{(2\pi)^{3/2}}\int d^3k e^{i\bm{k}\cdot\bm{x}}[h_{\bm{k}}(\eta)e_{ij}(\bm{k})+\tilde{h}_{\bm{k}}(\eta)\tilde{e}_{ij}(\bm{k})],
\end{equation}
where $e_{ij}(\bm{k})$ and $\tilde{e}_{ij}(\bm{k})$  are the plus and cross polarization tensors,
\begin{gather}
  e_{ij}(\bm{k})=\frac{1}{\sqrt{2}}\left[e_i(\bm{k})e_j(\bm{k})-\tilde{e}_i(\bm{k})\tilde{e}_j(\bm{k})\right], \\
  \tilde{e}_{ij}(\bm{k})=\frac{1}{\sqrt{2}}\left[e_i(\bm{k})\tilde{e}_j(\bm{k})+\tilde{e}_i(\bm{k})e_j(\bm{k})\right],
\end{gather}
and the orthonormal basis vectors $\bm e$ and  $\tilde{\bm e}$ are   orthogonal to $\bm{k}$, satisfying  $\bm e\cdot \tilde{\bm e}=\bm e \cdot \bm{k}= \tilde{\bm e}\cdot\bm{k}$.
The equation for the Fourier component of the tensor perturbation with either polarization induced by the scalar perturbation is
\cite{Ananda:2006af,Baumann:2007zm}
\begin{equation}
\label{eq:hk}
h''_{\bm{k}}+2\mathcal{H}h'_{\bm{k}}+k^2h_{\bm{k}}=4S_{\bm{k}},
\end{equation}
where a prime denotes the derivative with respect to the conformal time,  $h'_{\bm{k}}=dh_{\bm{k}}/d\eta$, $\mathcal{H}=a'/a $ is the conformal Hubble parameter,  the scalar source is
\begin{equation}
\label{hksource}
\begin{split}
S_{\bm{k}}=&\int \frac{d^3\tilde{k}}{(2\pi)^{3/2}}e_{ij}(\bm{k})\tilde{k}^i\tilde{k}^j
\left[2\Phi_{\tilde{\bm{k}}}\Phi_{\bm{k}-\tilde{\bm{k}}} \phantom{\frac{1}{2}}+ \right.\\
&\left.\frac{1}{\mathcal{H}^2} \left(\Phi'_{\tilde{\bm{k}}}+\mathcal{H}\Phi_{\tilde{\bm{k}}}\right)
\left(\Phi'_{\bm{k}-\tilde{\bm{k}}}+\mathcal{H}\Phi_{\bm{k}-\tilde{\bm{k}}}\right)\right],
\end{split}
\end{equation}
$\Phi_k$ is the Fourier component of the  Bardeen potential, and it can be related with its primordial value $\phi_{\bm{k}}$ by the transfer function
\begin{equation}
\Phi_{\bm{k}}=\Psi(k\eta)\phi_{\bm{k}}.
\end{equation}
The  transfer function $\Psi$ in the radiation domination is
\begin{equation}
\label{transfer}
\Psi(x)=\frac{9}{x^2}\left[\frac{\sin(x/\sqrt{3})}{x/\sqrt{3}}-\cos(x/\sqrt{3})\right].
\end{equation}
The primordial value $\phi_k$ relating to the primordial scalar power spectrum $\mathcal{P}_\zeta $ produced during inflation is
\begin{equation}
\label{phikeq4}
\langle\phi_{\bm{k}}\phi_{\tilde{\bm{k}}}\rangle
=\delta^{(3)}(\bm{k}+\tilde{\bm{k}})\frac{2\pi^2}{k^3}\left(\frac{3+3w}{5+3w}\right)^2 \mathcal{P}_\zeta(k),
\end{equation}
where $w$ is the equation of state  of matter in the Universe  determined by the time when the perturbations reenter the horizon.
The definition of the power spectrum $\mathcal{P}_h(k,\eta)$ for the SIGWs is
\begin{equation}
\label{eq:pwrh}
\langle h_{\bm{k}}(\eta)h_{\tilde{\bm{k}}}(\eta)\rangle
=\frac{2\pi^2}{k^3}\delta^{(3)}(\bm{k}+\tilde{\bm{k}})\mathcal{P}_h(k,\eta),
\end{equation}
and  the fractional energy density is
\begin{equation}
\label{density}
\Omega_{\mathrm{GW}}(k,\eta)=\frac{1}{24}\left(\frac{k}{aH}\right)^2\overline{\mathcal{P}_h(k,\eta)}.
\end{equation}
We solve the tensor equation \eqref{eq:hk} by the  Green function method, and the solution is
\begin{equation}\label{hk:green}
  h_k(\eta)=\frac{4}{a(\eta)}\int_{\eta_k}^{\eta}d \tilde{\eta}g_k(\eta,\tilde{\eta})a(\tilde{\eta})S_k(\tilde{\eta}),
\end{equation}
where the   corresponding  Green function is
\begin{equation}\label{green}
  g_k(\eta,\eta')=\frac{\sin\left[k(\eta-\eta')\right]}{k}.
\end{equation}
Substituting the solution \eqref{hk:green} into Eq. \eqref{eq:pwrh}, we  obtain the power spectrum for the SIGWs \cite{Baumann:2007zm,Ananda:2006af,Kohri:2018awv,Espinosa:2018eve}
\begin{equation}\label{ph}
\begin{split}
\mathcal{P}_h(k,\eta)=&
4\int_{0}^{\infty}dv\int_{|1-v|}^{1+v}du \left[\frac{4v^2-(1-u^2+v^2)^2}{4uv}\right]^2\\ &\times I_{RD}^2(u,v,x)\mathcal{P}_{\zeta}(kv)\mathcal{P}_{\zeta}(ku),
\end{split}
\end{equation}
Using the definition of  the fractional energy density, we obtain
\begin{equation}
\label{SIGWs:gwres1}
\begin{split}
\Omega_{\mathrm{GW}}(k,\eta)=&\frac{1}{6}\left(\frac{k}{aH}\right)^2\int_{0}^{\infty}dv\int_{|1-v|}^{1+v}du \\
&\times\left[\frac{4v^2-(1-u^2+v^2)^2}{4uv}\right]^2\\
&\times\overline{I_{\text{RD}}^{2}(u, v, x)} \mathcal{P}_{\zeta}(kv)\mathcal{P}_{\zeta}(ku),
\end{split}
\end{equation}
where $u=|\bm{k}-\tilde{\bm{k}}|/k$, $v=\tilde{k}/k$, $x=k\eta$  and the integral kernel $I_{\text{RD}}$  is \cite{Espinosa:2018eve,Lu:2019sti}
\begin{equation}
\label{irdeq1}
\begin{split}
I_{\text{RD}}(u, v, x)=&\int_1^x dy\, y \sin(x-y)\{3\Psi(uy)\Psi(vy)\\
&+y[\Psi(vy)u\Psi'(uy)+v\Psi'(vy)\Psi(uy)]\\
&+y^2 u v \Psi'(uy)\Psi'(vy)\}.
\end{split}
\end{equation}
This integral kernel  can split into two parts, one is the sine part and the other is the cosine part,
\begin{equation}
\label{s:irdeq1}
I_{\text{RD}}(u, v, x)=\frac{1}{9x}\left(I_s\sin{x}+I_c\cos{x}\right),
\end{equation}
where the definitions of $I_c$ and $I_s$ are
\begin{equation}
\label{iceq11}
\begin{split}
    I_{c}(u, v, x)=&-4\int_1^x y \sin(y) f(y) d y\\
    =&T_{c}(u, v, x)-T_{c}(u, v, 1),
\end{split}
\end{equation}
\begin{equation}
\label{iseq11}
\begin{split}
    I_{s}(u, v, x)=&4 \int_1^x y \cos(y) f(y) d y\\
    =&T_{s}(u, v, x)-T_{s}(u, v, 1),
\end{split}
\end{equation}
and
\begin{equation}
\label{tceq1}
T_c(u, v, x)=-4\int_0^x y \sin(y) f(u,v,y) d y,
\end{equation}
\begin{equation}
\label{tseq1}
T_s(u, v, x)=4 \int_0^x y \cos(y) f(u,v,y) d y,
\end{equation}
with
\begin{equation}
\label{irdeq1a}
\begin{split}
f(u,v,x)=&2 \Psi(v x) \Psi(u x)+\\
&\left[\Psi(v x)+vx\Psi'(v x)\right]\left[\Psi(u x)+ux\Psi'(u x)\right].
\end{split}
\end{equation}
By using the definition of the transfer function \eqref{transfer},  Eq. \eqref{tceq1}   becomes
\begin{equation}
\label{tceq2}
\begin{split}
    T_c=&\frac{-27}{8u^3v^3x^4}\left[\vphantom{\frac12}-48uvx^2(x\cos x+3\sin x)\cos\frac{ux}{\sqrt{3}}\cos\frac{vx}{\sqrt{3}}\right.\\
    &+48\sqrt{3} x^2\cos{x}\left(v\cos{\frac{vx}{\sqrt{3}}}\sin{\frac{ux}{\sqrt{3}}}
    +u\cos{\frac{ux}{\sqrt{3}}}\sin{\frac{vx}{\sqrt{3}}}\right)\\
    &+8\sqrt{3} x\sin{x}\left(\vphantom{\frac12}
    [18-x^2(u^2+3-v^2)]v\cos{\frac{vx}{\sqrt{3}}}\sin{\frac{ux}{\sqrt{3}}}\right.\\
    &\left.+[18-x^2(v^2+3-u^2)]u
    \cos{\frac{ux}{\sqrt{3}}}\sin{\frac{vx}{\sqrt{3}}}\right)\\
    &+24x[-6+x^2(3-u^2-v^2)]\cos{x}
    \sin{\frac{ux}{\sqrt{3}}}\sin{\frac{vx}{\sqrt{3}}}\\
    &\left.+24[-18+x^2(3+u^2+v^2)]
    \sin{x}\sin{\frac{ux}{\sqrt{3}}}\sin{\frac{vx}{\sqrt{3}}}\right]\\
    &-\frac{27(u^2+v^2-3)^2}{4u^3v^3}\left(\text{Si}
    \left[\left(1-\frac{u-v}{\sqrt{3}}\right)x\right]\right.\\
    &+\text{Si}\left[\left(1+\frac{u-v}{\sqrt{3}}\right)x\right]
    -\text{Si}\left[\left(1-\frac{u+v}{\sqrt{3}}\right)x\right]\\
    &\left.-\text{Si}\left[\left(1+\frac{u+v}{\sqrt{3}}\right)x\right]\right),
\end{split}
\end{equation}
and Eq. \eqref{tseq1} becomes
\begin{equation}
\label{tseq2}
\begin{split}
    T_s=&\frac{27}{8u^3v^3x^4}\left[\vphantom{\frac12}
    48uvx^2(x\sin x-3\cos x)\cos\frac{ux}{\sqrt{3}}\cos\frac{vx}{\sqrt{3}}\right.\\
    &-48\sqrt{3} x^2\sin{x}\left(v\cos{\frac{vx}{\sqrt{3}}}\sin{\frac{ux}{\sqrt{3}}}+u\cos{\frac{ux}{\sqrt{3}}}\sin{\frac{vx}{\sqrt{3}}}\right)\\
    &+8\sqrt{3} x\cos{x}\left(\vphantom{\frac12}\right.[18-x^2(u^2+3-v^2)]v\cos{\frac{vx}{\sqrt{3}}}\sin{\frac{ux}{\sqrt{3}}}\\
    &+[18-x^2(v^2+3-u^2)]u\cos{\frac{ux}{\sqrt{3}}}\sin{\frac{vx}{\sqrt{3}}}\left.\vphantom{\frac12}\right)\\
    &+24x[6-x^2(3-u^2-v^2)]\sin{x}\sin{\frac{ux}{\sqrt{3}}}\sin{\frac{vx}{\sqrt{3}}}\\
    &\left.+24[-18+x^2(3+u^2+v^2)]\cos{x}\sin{\frac{ux}{\sqrt{3}}}\sin{\frac{vx}{\sqrt{3}}}\right]\\
    &-\frac{27(u^2+v^2-3)}{u^2v^2}+\frac{27(u^2+v^2-3)^2}{4u^3 v^3}\\
    &\times\left(\text{Ci}\left[\left(1-\frac{u-v}{\sqrt{3}}\right)x\right]
    +\text{Ci}\left[\left(1+\frac{u-v}{\sqrt{3}}\right)x\right]\right.\\
    &\-\text{Ci}\left[\left|1-\frac{u+v}{\sqrt{3}}\right|x\right]-
    \text{Ci}\left[\left(1+\frac{u+v}{\sqrt{3}}\right)x\right]\\
    &\left.+\ln\left|\frac{3-(u+v)^2}{3-(u-v)^2}\right|\right).
\end{split}
\end{equation}
where the functions  $\text{Si}(x)$ and  $\text{Ci}(x)$  are the sine-integral function and cosine-integral function defined as
\begin{equation}
    \text{Si}(x)=\int_0^x d y\frac{\sin y}{y},\quad \text{Ci}(x)=-\int_x^\infty d y \frac{\cos y}{y}.
\end{equation}
At late times, the modes are deeply  inside the horizon during radiation dominated era,  $x\ll1 $, the integral kernel becomes
\begin{equation}
\label{hijkeq16}
\begin{split}
    I_{\text{RD}}(u,v,x\rightarrow\infty)=-\frac{3\pi(u^2+v^2-3)^2\Theta(u+v-\sqrt{3})}{4u^3v^3x}\cos{x}\\
    -\frac{1}{9x}\left(T_c(u,v,1)\cos{x}+\tilde{T}_s(u,v,1)\sin{x}\right),
\end{split}
\end{equation}
where
\begin{equation}
\label{hijkeq16a}
\begin{split}
\tilde{T}_s(u,v,1)=&T_s(u,v,1)+\frac{27(u^2+v^2-3)}{u^2v^2}\\
&-\frac{27(u^2+v^2-3)^2}{4u^3 v^3}\times\\
&\ln\left|\frac{3-(u+v)^2}{3-(u-v)^2}\right|.
\end{split}
\end{equation}
So the time average of the integral kernel  is \cite{Lu:2019sti}
\begin{equation}
\label{averaged}
\begin{split}
    \overline{I^2_{\text{RD}}(u,v,x\rightarrow\infty)}=\frac{1}{2x^2}\left[\left(\frac{\tilde{T}_s(u,v,1)}{9}\right)^2+
    \left(\frac{T_c(u,v,1)}{9}+\right.\right.\\
    \left.\left.\frac{3\pi(u^2+v^2-3)^2\Theta(u+v-\sqrt{3})}{4u^3v^3}\right)^2
    \right].
\end{split}
\end{equation}
Substituting the above equation into  Eq. \eqref{SIGWs:gwres1}, we obtain the energy density of the SIGWs during the radiation dominated era.


%
 
\end{document}